\documentclass[a4paper,12pt]{article}
\usepackage{graphicx}
\usepackage{subfig}
\title{Precision measurement of Higgs decay branching ratios to bottom quarks and gluons at the ILC}
\author{Y. Banda, T. Lastovicka, A. Nomerotski \\Particle Physics Department, University of Oxford \\Denys Wilkinson Building, Keble Road \\Oxford, OX1 3RH}
\date{}
\begin{document}
\maketitle
\begin{abstract}
The measurement of hadronic Higgs Boson branching ratios H$\rightarrow b\bar{b}$, H$\rightarrow gg$ for a light Standard Model-like Higgs boson produced at 250 GeV centre of mass energy at the International Linear Collider (ILC) is presented. The tools and techniques used for the analysis are briefly discussed.
\end{abstract}

\section{Introduction} 

\begin{figure}[htbp]
\begin{center}
\includegraphics[scale=0.35]{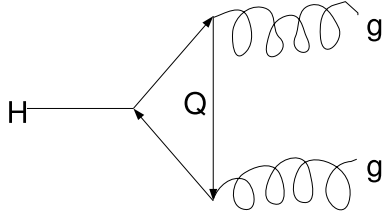}
\end{center}
\caption{Higgs decay to gluons mediated by a heavy quark loop.}
\label{Hggs}
\end{figure}

The measurement of Higgs boson branching ratios is one of the main features of the International Linear Collider (ILC) program~\cite{ilcrdr}.
For Higgs masses below 140 GeV, hadronic branching ratios can be precisely measured at the ILC. The final states have significant rates and 
the micro-vertex detector allows for good flavour identification. This measurement of relative couplings of the Higgs boson to fermions will allow 
to confirm the prediction of the Higgs mechanism that they are proportional to fermion masses. The branching fraction to b and c quarks is larger than 
the branching fraction to light quarks due to the mass. The Higgs decay to gluons in the Standard Model is mediated by heavy quark loops as shown in Figure~\ref{Hggs}. The branching ratio to gluons is indirectly related to $t\bar{t}$H Yukawa coupling~\cite{yukawa} and would probe the existence of new strongly interacting particles that couple to the Higgs and are too heavy to be produced directly.

In this study we consider Higgs decays to bottom quarks and gluons in two- and four-jet configurations. For a $m_H$ = 120 GeV SM Higgs boson, we expect BR(H $\rightarrow b\bar{b}$) = 67.92\% and BR(H $\rightarrow gg$) = 7.06\%~\cite{hdecay}. These decay modes exercise the tagging of heavy and light quarks~\cite{yb,zh}.  

\section{Analysis}

The analysis strategy closely follows the one implemented in ~\cite{yb}.

\subsection{Event Generation and Detector Simulation}

In this study, the signal sample includes a Higgs boson produced in the Higgs-strahlung process, $e^{+}e^{-} {\to }$ ZH. The mass of the
Higgs is assumed to be 120 GeV. Standard Model events (mainly WW, ZZ and qq pairs) and Higgs decays to other fermions other than gluons/bottom quarks are considered as backgrounds. Signal and background events are produced at the centre of mass energy $\sqrt{s}=250$~GeV, total integrated luminosity of 250~fb$^{-1}$ and $\mp$80\% electron and $\pm$30\% positron polarization~\cite{yb}. The choice of energy in this analysis maximizes the cross-section value for Higgs-strahlung. All 0, 2 and 4 fermion final states were generated with the Whizard Monte Carlo Event Generator~\cite{whiz}. 
PYTHIA~\cite{pyth} was used for the final state QED and QCD parton showering, fragmentation and decay to provide final-state 
observable particles. Photons from beamstrahlung and initial state radiation were also included in the simulations.

Geant4 toolkit~\cite{geant,geant2} was used to simulate detector response to generated events and SLIC~\cite{slic} provided access to the Monte Carlo events, the detector geometry and the output of the detector hits.

\subsection{Event Selection}

The event selection is identical to that done in~\cite{yb} where further analysis details can be found. There are two channels studied in this analysis: the 2-jet neutrino channel (when the Z boson decays to neutrinos and the Higgs decays to gg/bb) and the 4-jet hadronic channel (Z decays to quark pairs and Higgs to gg/bb). The selection of events defining each channel is based on the visible energy and the number of leptons in the event. Hadronic jet reconstruction, in which events are forced into two or four jet configurations, is achieved by the DURHAM algorithm~\cite{durham}. Identification of primary, secondary and tertiary vertices is performed 
by the topological vertex finder ZVTOP which is part of the vertexing package developed by the LCFI collaboration~\cite{lcfi}.  

After the channel classification, a neural network analysis is performed in order to discriminate the signal and background events. The 
surviving events are then used for the branching ratio calculation.

\subsubsection{Neutrino Channel}

The signal in this channel consists of two jets from the Higgs recoiling against two neutrinos from the Z boson decay. The missing mass is expected to
be consistent with the Z mass and the invariant mass of the 2 jets consistent with the Higgs mass. The main backgrounds in this channel include WW pairs where
\begin{itemize}
\item one W decays hadronically. 
\item and the other W decays into a neutrino and lepton which escapes undetected along the beampipe,
\end{itemize}
ZZ pairs in which one Z decays hadronically and the other into neutrinos and qq pairs.

For this channel, no leptons are accepted and the visible energy is required to be between 90 and 160 GeV. Leptons are defined as electrons or muons with minimum 15 GeV momentum. The following pre-selection cuts are applied to reduce the background:
\begin{enumerate}
\item 20 $<$ P$_T$ (transverse momentum) of jet $<$ 90 GeV. Most SM background events are softer than signal events.
\item $n_{tracks}$ $>$ 4. More than 4 charged tracks for leptonic event rejection.
\item $-\log(y_{min}$) $<$ 0.8. Durham algorithm parameter which determines number of jets in events. It is used to reject fully hadronic WW and 
ZZ events.
\item thrust $< 0.95$. Background events are more boosted and less spherical than signal events.
\item $\cos(\theta_{thrust})$$< 0.98$. Signal events occur more centrally in the detector than background events.
\item $100^{\circ}$ $<$ angle between jets $< 170^{\circ}$.
\item 100 GeV $<$ di-jet invariant mass $<$ 140 GeV. The Higgs mass is expected to be 120 GeV.  
\item Highest reconstructed photon energy $<$ 10 GeV. Required to reject 2-fermion events with large ISR.
\end{enumerate}

Tables~\ref{tab:bnuevts} and ~\ref{tab:gnuevts} show the number of events before and after pre-selection cuts for $b\bar{b}$ and gg in the neutrino channel.
\begin{table}[htpb]
\centering
\begin{tabular}{|c|c|c|c|}
\hline
Cuts & SM background & Higgs background & Signal \\ \hline
(i) Before Classification  & 9275594683 & 6048 & 12187 \\ \hline
(0) After Classification & 45936973 & 5248 & 11580 \\ \hline 
(1) & 18374789 & 5053 & 11243 \\ \hline
(2) & 17123140 & 4255 & 10609 \\ \hline
(3) & 6849256 & 3976 & 10196\\ \hline
(4) & 685329 & 3782 & 8427 \\ \hline
(5) & 627113 & 3562 & 8027 \\ \hline
(6) & 576422 & 3403 & 7907 \\ \hline
(7) & 203292 & 2786 & 6801 \\ \hline
(8) & 109057 & 2737 & 6707 \\ \hline
\end{tabular}
\caption{Number of $b\bar{b}$ events before channel classification, after channel classification and after pre-selection cuts in the neutrino mode.}
\label{tab:bnuevts}
\end{table}

\begin{table}[htpb]
\centering
\begin{tabular}{|c|c|c|c|}
\hline
Cuts & SM background & Higgs background & Signal \\ \hline
(i) Before Classification  & 9275594683 & 17547 & 992 \\ \hline
(0) After Classification & 45936973 & 15820 & 986 \\ \hline 
(1) & 18374789 & 15317 & 979 \\ \hline
(2) & 17123140 & 13896 & 968 \\ \hline
(3) & 6849256 & 13252 & 920 \\ \hline
(4) & 685329 & 11343 & 865 \\ \hline
(5) & 627113 & 10766 & 823 \\ \hline
(6) & 576422 & 10515 & 795 \\ \hline
(7) & 203292 & 8818 & 769 \\ \hline
(8) & 109057 & 8684 & 759 \\ \hline
\end{tabular}
\caption{Number of gg events before channel classification, after channel classification and after pre-selection cuts in the neutrino mode.}
\label{tab:gnuevts}
\end{table}

\begin{figure}[htbp]
\begin{center}
\subfloat[]{\includegraphics[scale=0.35]{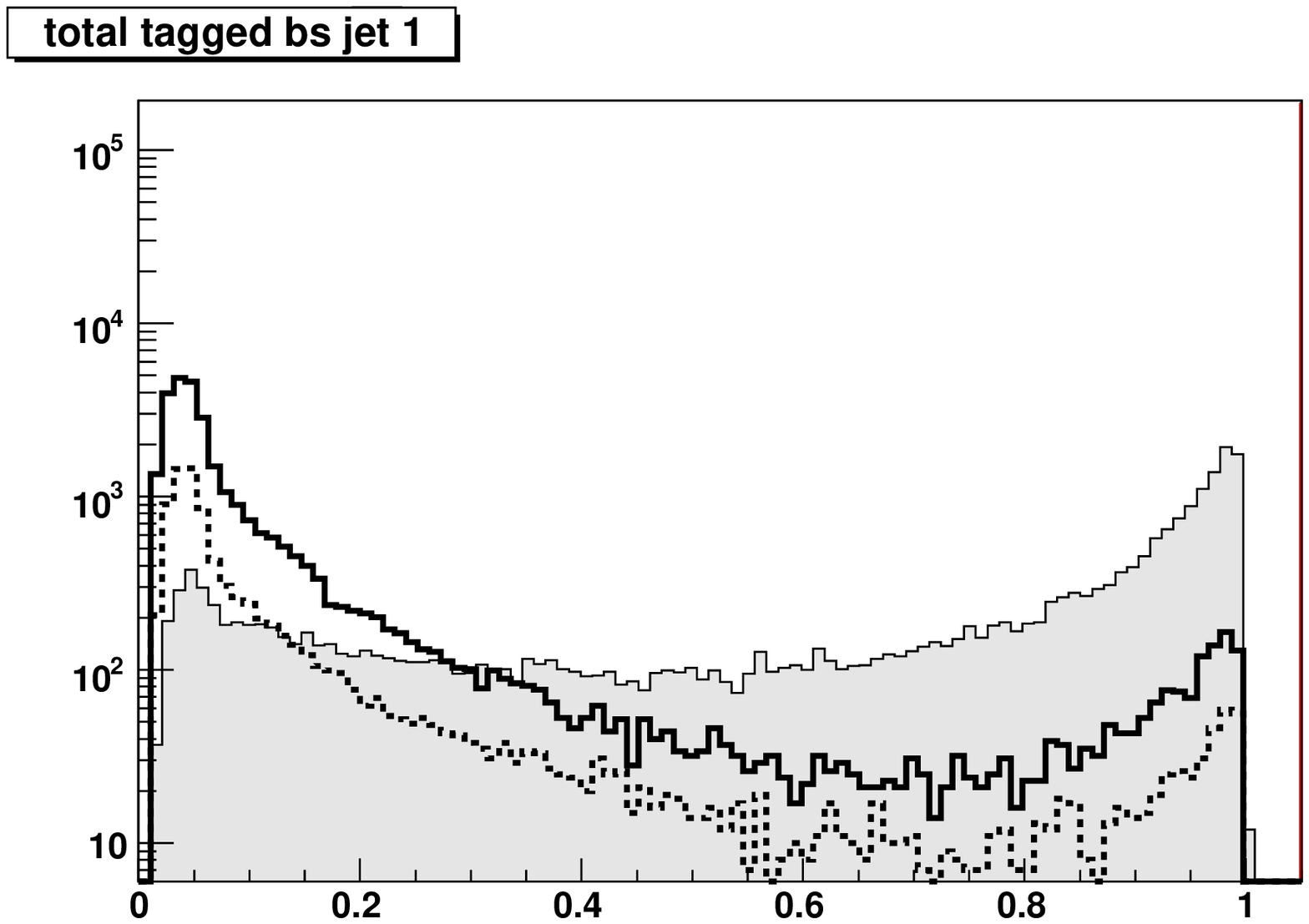}}
\subfloat[]{\includegraphics[scale=0.35]{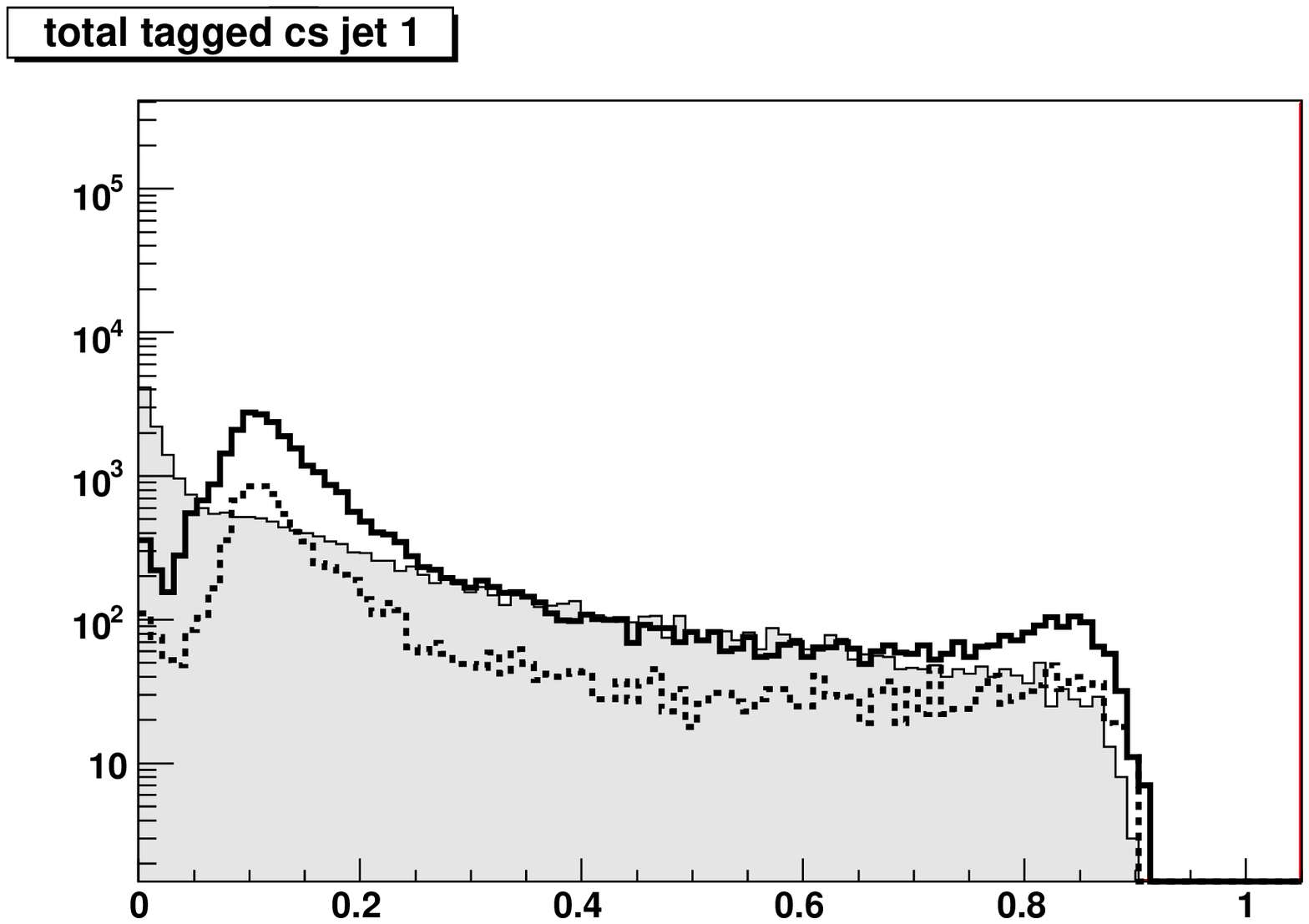}} \\
\subfloat[]{\includegraphics[scale=0.35]{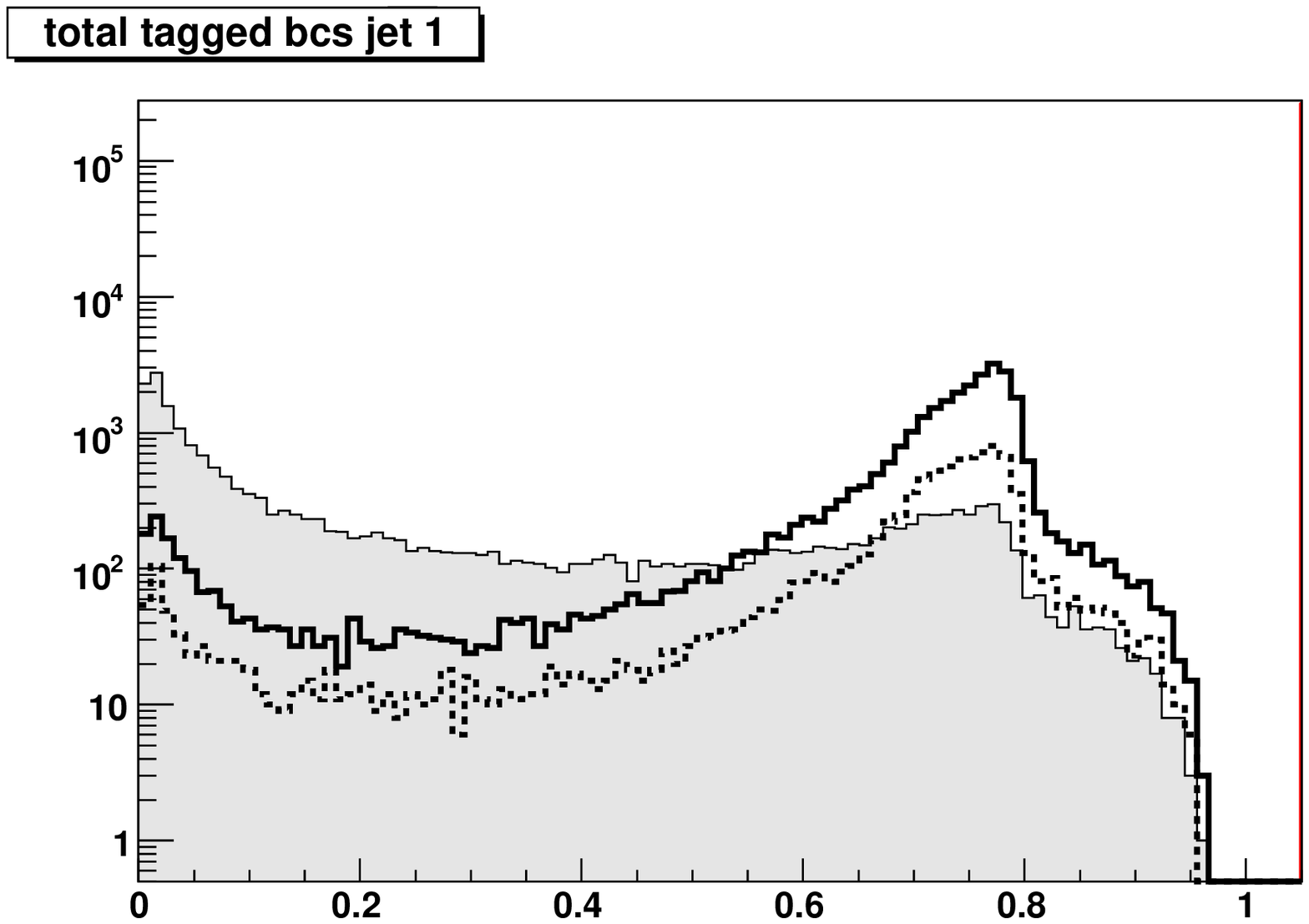}} 
\end{center}
\caption{Distributions of flavour tagging variables for $b\bar{b}$ leading jet in the neutrino channel: (a) b-tag; (b) c-tag; (c) c-tag with b background only. Solid curves are SM background, dashed curves are Higgs background sample and filled histograms are the signal.}
\label{bbnn}
\end{figure}

\begin{figure}[htbp]
\begin{center}
\subfloat[]{\includegraphics[scale=0.35]{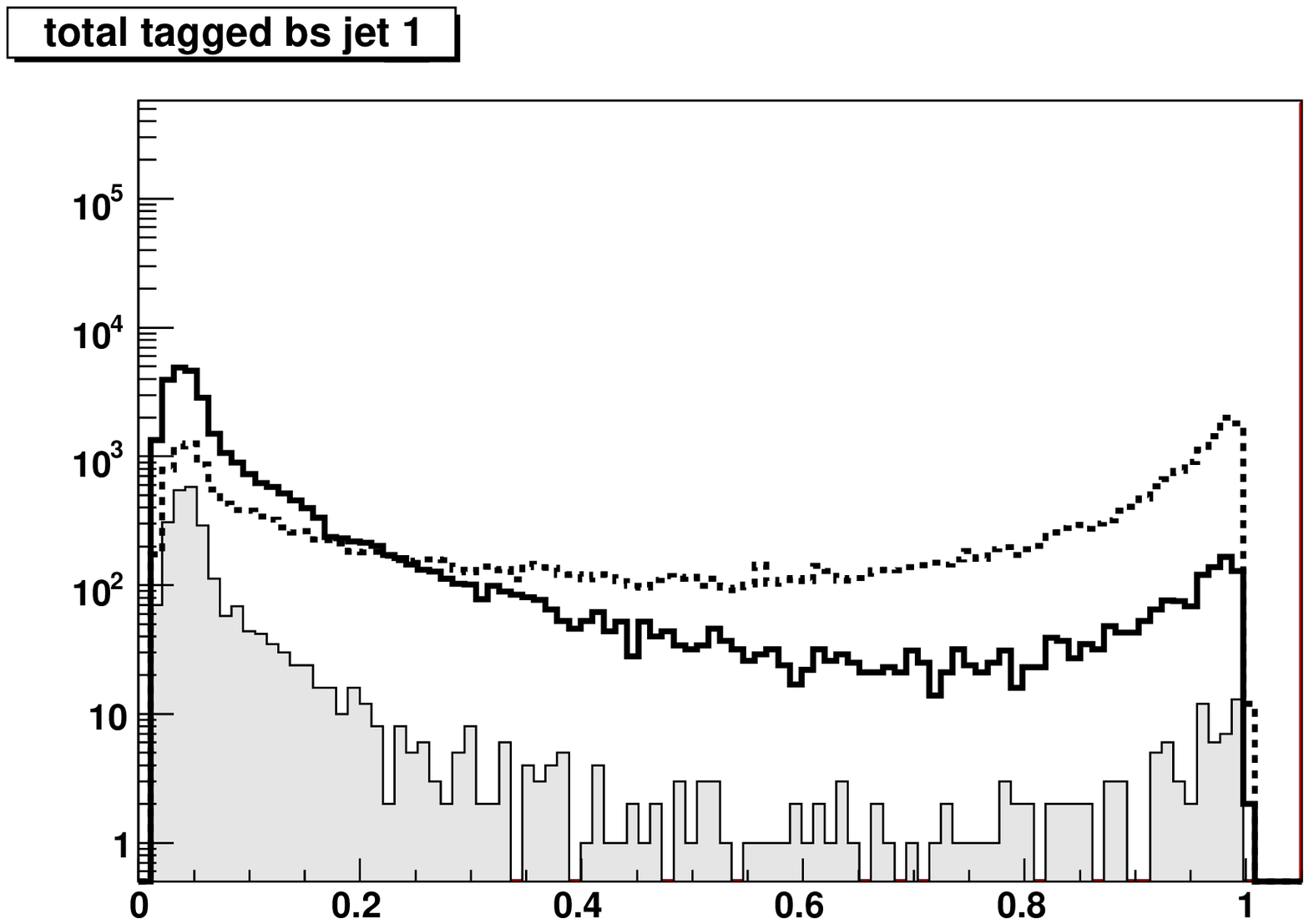}}
\subfloat[]{\includegraphics[scale=0.35]{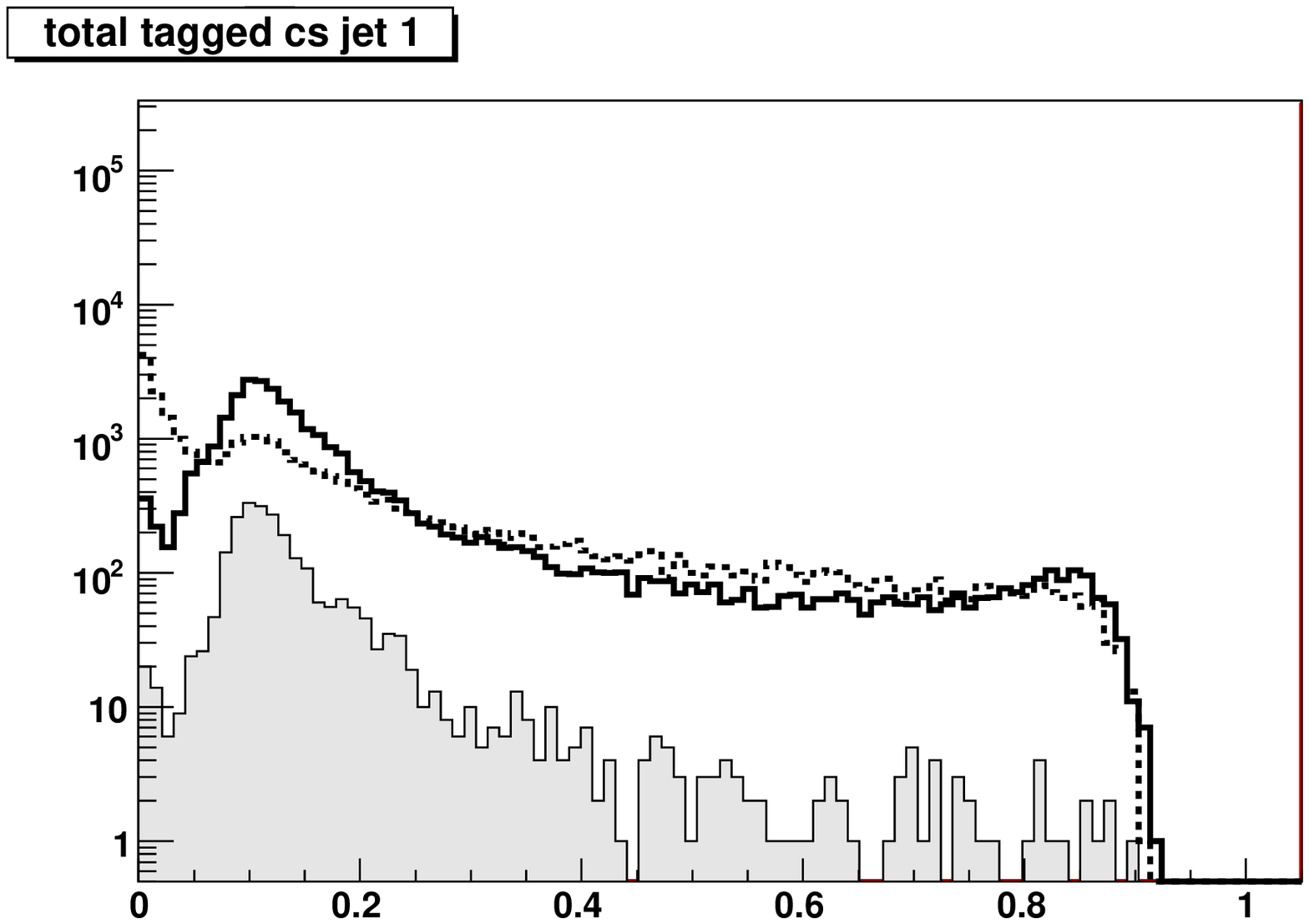}} \\
\subfloat[]{\includegraphics[scale=0.35]{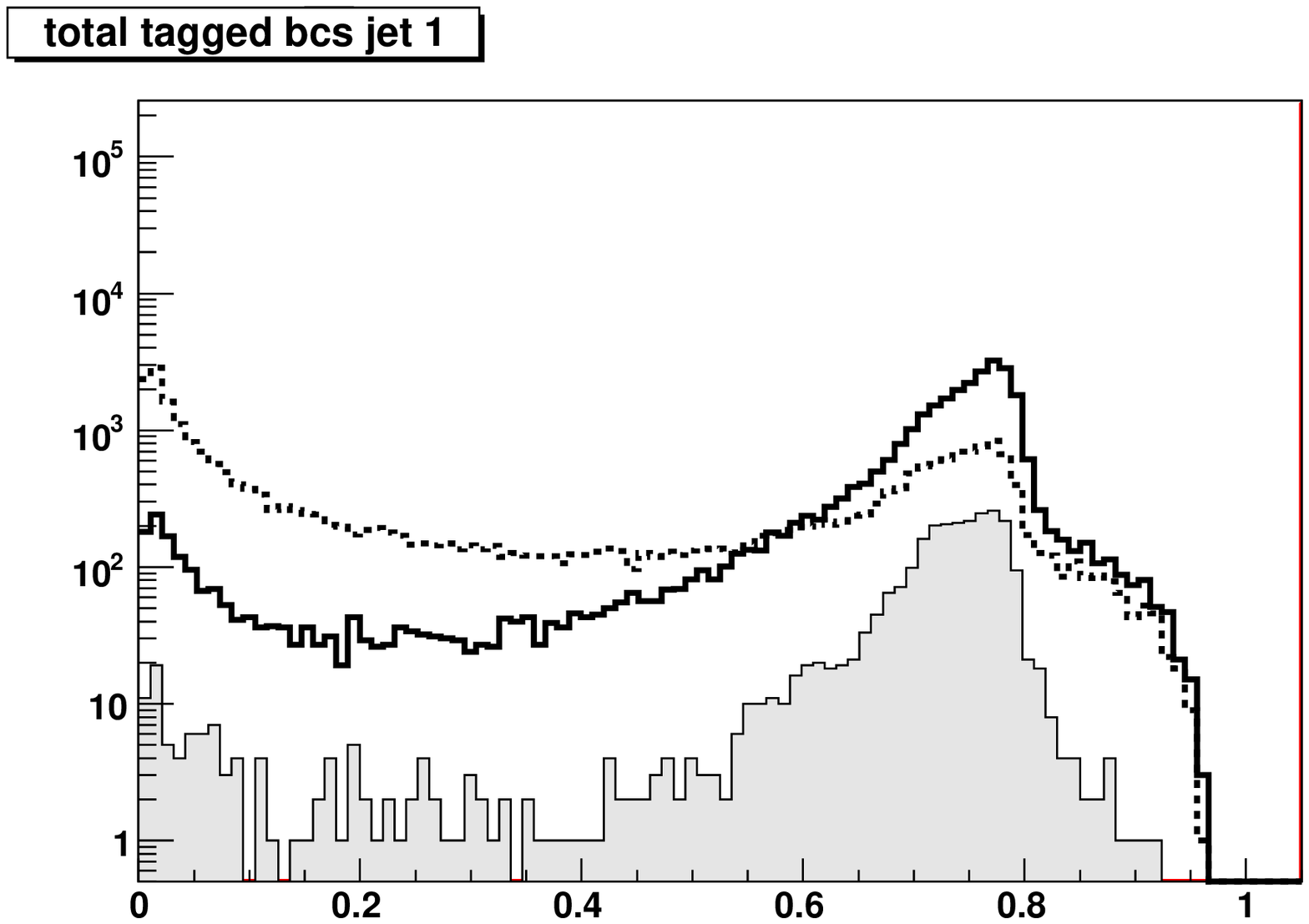}} 
\end{center}
\caption{Distributions of flavour tagging variables for gg leading jet in neutrino channel: (a) b-tag; (b) c-tag; (c) c-tag with b background only. Solid curves are SM background, dashed curves are Higgs background sample and filled histograms are the signal.}
\label{ggnn}
\end{figure}

Events that survive the pre-selection cuts undergo a neural network (NN) analysis to help discriminate the signal and background. The input variables
to the neural network include all pre-selection variables and also the jet flavour tagging outputs produced using the LCFI package. There are three
types of jet flavour tag outputs: b-tag, c-tag and c-tag with b background only. Figure~\ref{bbnn} and Figure~\ref{ggnn} show distributions of the three LCFI flavour tags, `b-tag', `c-tag' and  `c-tag with b background only' for the leading b and gluon jets. For the bb/gg scenarios the signal is defined as only H$\rightarrow$ bb/gg events and the Higgs background is all other Higgs decays other than H$\rightarrow$ bb/gg. All histograms are normalized to 250~fb$^{-1}$.

The tagging of jets works well. B-jets have relatively long lifetimes and decay topology with up to two secondary vertices and this makes them very distinct. Gluonic jets are tagged as light quark jets.

\subsubsection{Hadronic Channel}

In the hadronic channel both the Higgs and Z bosons decay into two partons (quarks or gluons) resulting in a 4-jet configuration. For the signal, the invariant mass of two jets is required to be consistent with the Higgs mass and the other two jets should have a mass corresponding to the Z boson.
To reduce combinatorial effects,  kinematic fitting~\cite{yb,kinfit} is performed to identify jets from the Higgs boson and jets from the Z boson.
Jet pairing is performed before kinematic fitting. For the 4-jet events we have 6 possible pairings of the jets and 3 possible associations of the 4 jets to the Z and H bosons. For the 6 possible pairings we calculate the invariant mass of each pair and compare with the masses of bosons. For each event we calculate 
\begin{equation}
d = (m_{ij}-m_{Z})^2 + (m_{kl}-m_{H})^2
\end{equation}
The pairing that minimizes {\it d} is chosen. The signal four-jet configuration has, on average, the leading and third jets coming from the Higgs 
boson and the second and last jets coming from the Z boson.

The kinematic constraints used for the fitter are the jet four-momenta, centre of mass energy (250 GeV) and the invariant mass difference of two jet pairs.

Classification of the hadronic channel requires that no leptons are present in the event and a minimum of 170 GeV of visible energy. Apart from the variables used in the neutrino channel, the other variables/cuts applied in the hadronic channel are the angle between Z boson jets and the invariant mass of jets coming from the Z boson. Table~\ref{tab:HZ2} shows the selection cuts in the hadronic channel.
\begin{table}[htpb]
\centering
\begin{tabular}{lllll}
\hline Cuts & && selection & value  \\
\hline
1. & && number of charged tracks per jet & $>$ 4\\
2. & && $-\log(y_{min}$) & $<$ 2.7 \\
3. & && thrust & $< 0.95$ \\
4. & && $ \cos(\theta_{thrust})$ & $< 0.96$ \\
5. & $105^{\circ}$ &$<$ & angle between jet 1 and 3 &$< 165^{\circ}$ \\
6. & $70^{\circ}$ &$<$ & angle between jet 2 and 4 &$< 160^{\circ}$ \\
7. & 110 GeV &$<$ & invariant mass of Higgs candidate after fit &$<$ 140 GeV \\
8. & 80 GeV &$<$ & invariant mass of Z candidate after fit &$<$ 110 GeV \\
9. & && Highest reconstructed photon energy & $<$ 10 GeV \\
\hline
\end{tabular}
\caption{Selections for the four-jet analysis.}
\label{tab:HZ2}
\end{table}

Tables~\ref{tab:4bnuevts} and ~\ref{tab:4gnuevts} show the number of events before and after pre-selection cuts for $b\bar{b}$ and gg in the neutrino channel.
\begin{table}[htpb]
\centering
\begin{tabular}{|c|c|c|c|}
\hline
Cuts & SM background & Higgs background & Signal \\ \hline
(i) Before Classification  & 9275594683 & 17816 & 35629 \\ \hline
(0) After Classification & 39398366 & 11875 & 30985 \\ \hline 
(1) & 18601753 & 8821 & 24398 \\ \hline
(2) & 13921271 & 7582 & 20736 \\ \hline
(3) & 8737017 & 6472 & 18391 \\ \hline
(4) & 7943851 & 5977 & 17130 \\ \hline
(5) & 5871237 & 5681 & 16339 \\ \hline
(6) & 4898312 & 5678 & 16326 \\ \hline
(7) & 1917231 & 5633 & 16108 \\ \hline
(8) & 1561432 & 5622 & 16108 \\ \hline
(9) & 967312 & 5486 & 15805 \\ \hline
\end{tabular}
\caption{Number of $b\bar{b}$ events before channel classification, after channel classification and after pre-selection cuts in the hadronic mode.}
\label{tab:4bnuevts}
\end{table}

\begin{table}[htpb]
\centering
\begin{tabular}{|c|c|c|c|}
\hline
Cuts & SM background & Higgs background & Signal \\ \hline
(i) Before Classification  & 9275594683 & 50453 & 2992 \\ \hline
(0) After Classification & 39398366 & 39894 & 2965 \\ \hline 
(1) & 18601753 & 30717 & 2503 \\ \hline
(2) & 13921271 & 26119 & 2199 \\ \hline
(3) & 8737017 & 22977 & 1887 \\ \hline
(4) & 7943851 & 21396 & 1738 \\ \hline
(5) & 5871237 & 20373 & 1648 \\ \hline
(6) & 4898312 & 20357 & 1648 \\ \hline
(7) & 1917231 & 20103 & 1638 \\ \hline
(8) & 1561432 & 20092 & 1638 \\ \hline
(9) & 967312 & 19860 & 1611 \\ \hline
\end{tabular}
\caption{Number of gg events before channel classification, after channel classification and after pre-selection cuts in the hadronic mode.}
\label{tab:4gnuevts}
\end{table}

\begin{figure}[htbp]
\begin{center}
\subfloat[]{\includegraphics[scale=0.35]{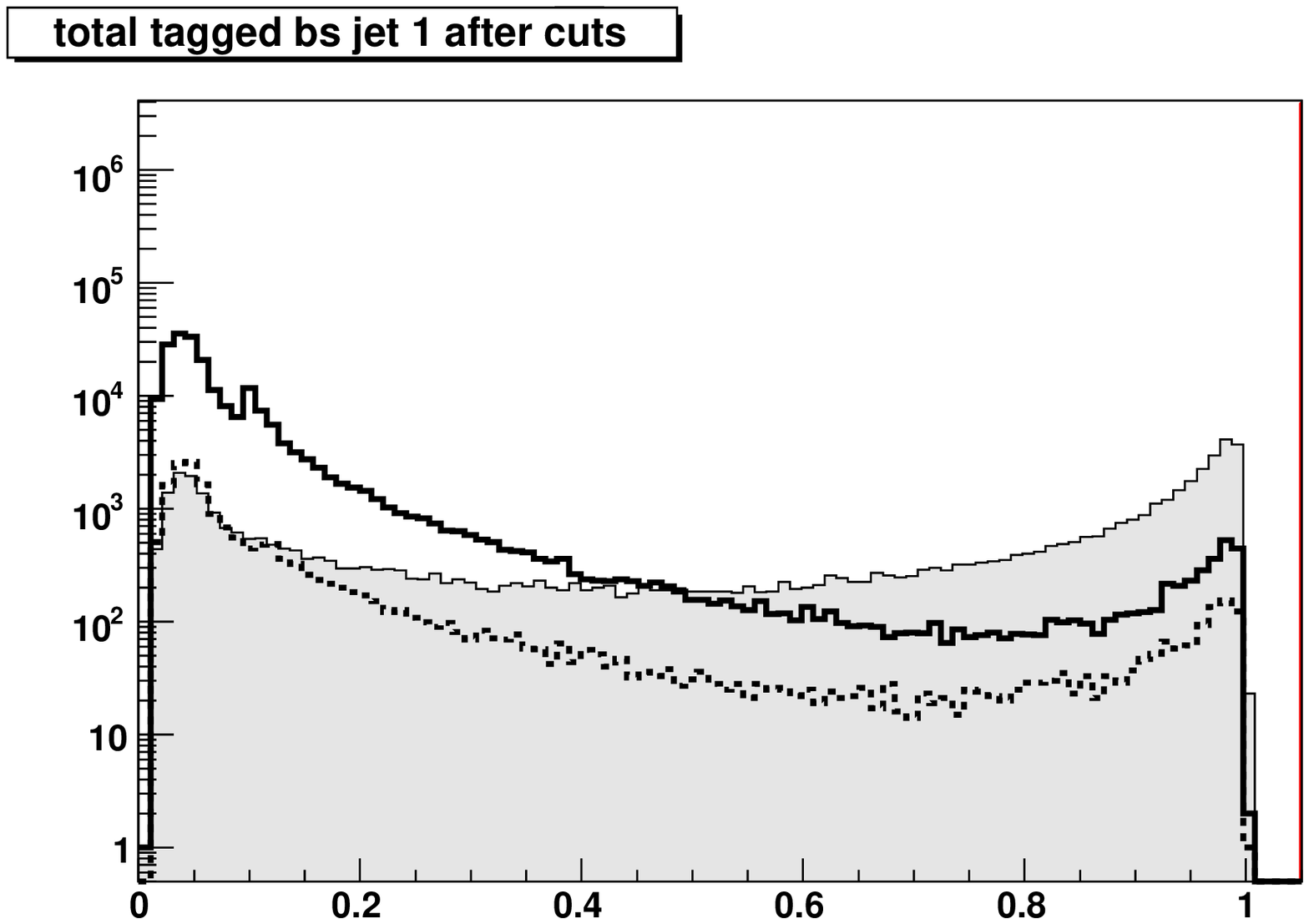}}
\subfloat[]{\includegraphics[scale=0.35]{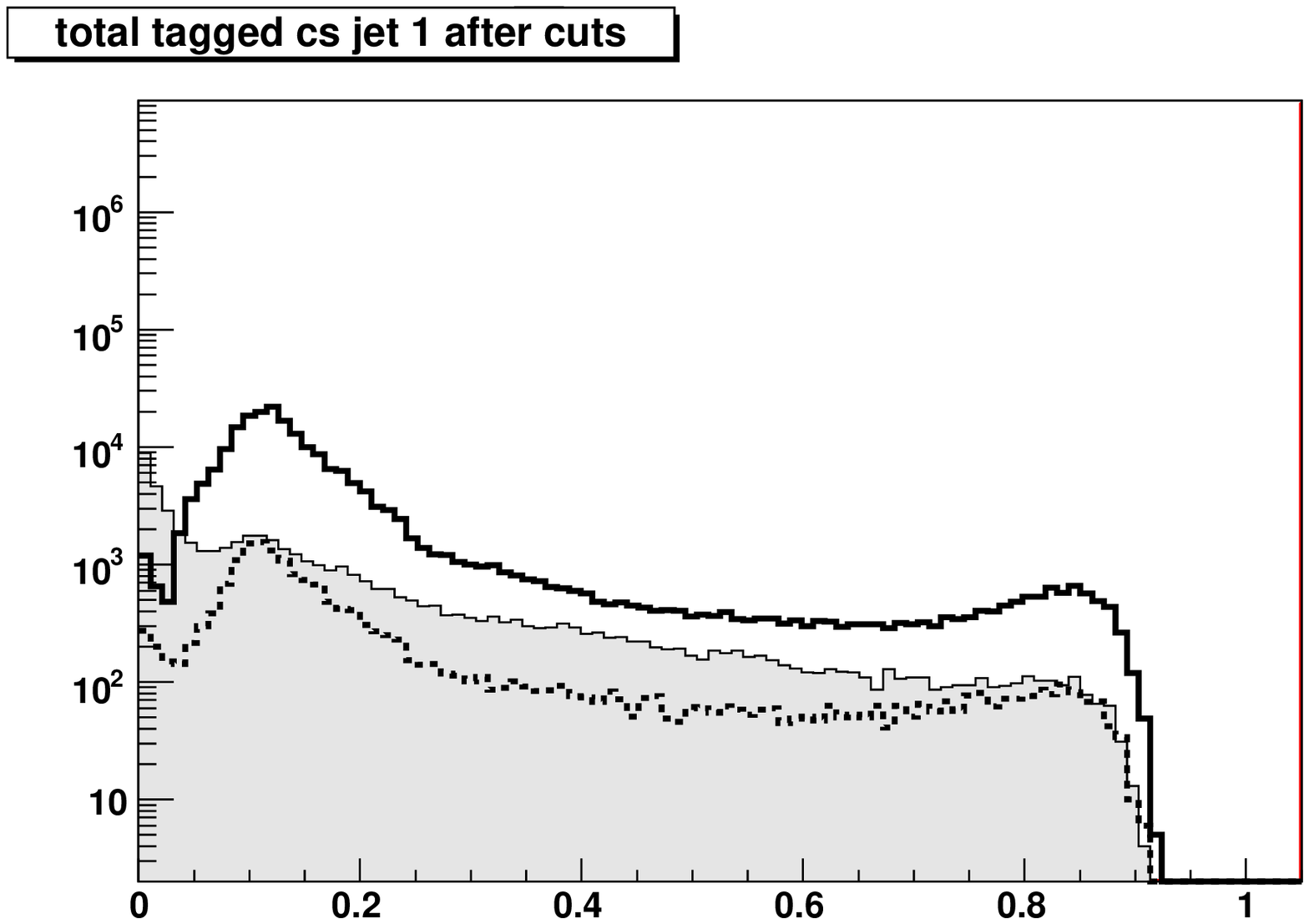}} \\
\subfloat[]{\includegraphics[scale=0.35]{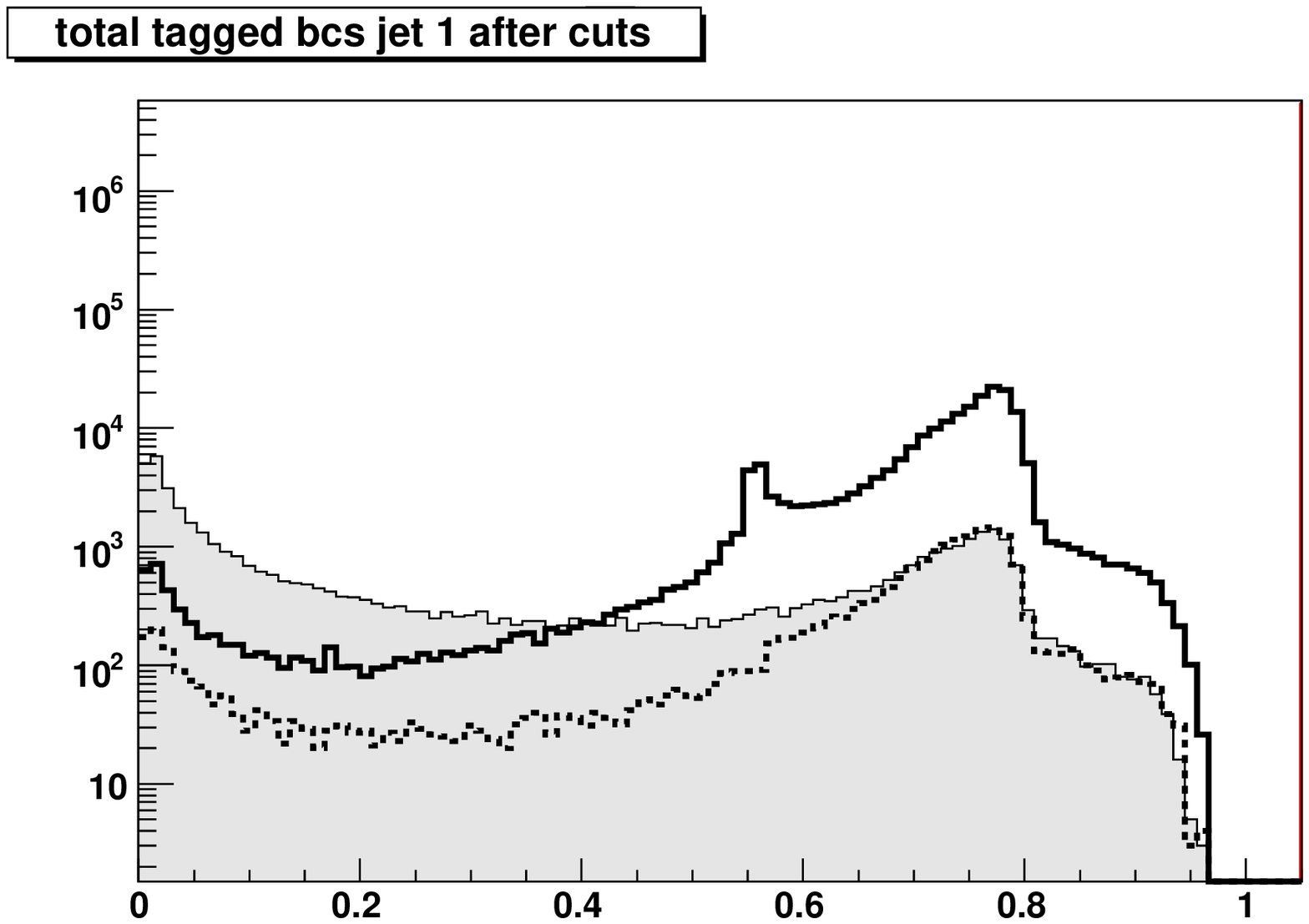}} 
\end{center}
\caption{Distributions of flavour tagging variables for $b\bar{b}$ in hadronic channel: Leading jet (a) b-tag; (b) c-tag; (c) c-tag with b background only.}
\label{4bnn1}
\end{figure}

\begin{figure}[htbp]
\begin{center}
\subfloat[]{\includegraphics[scale=0.35]{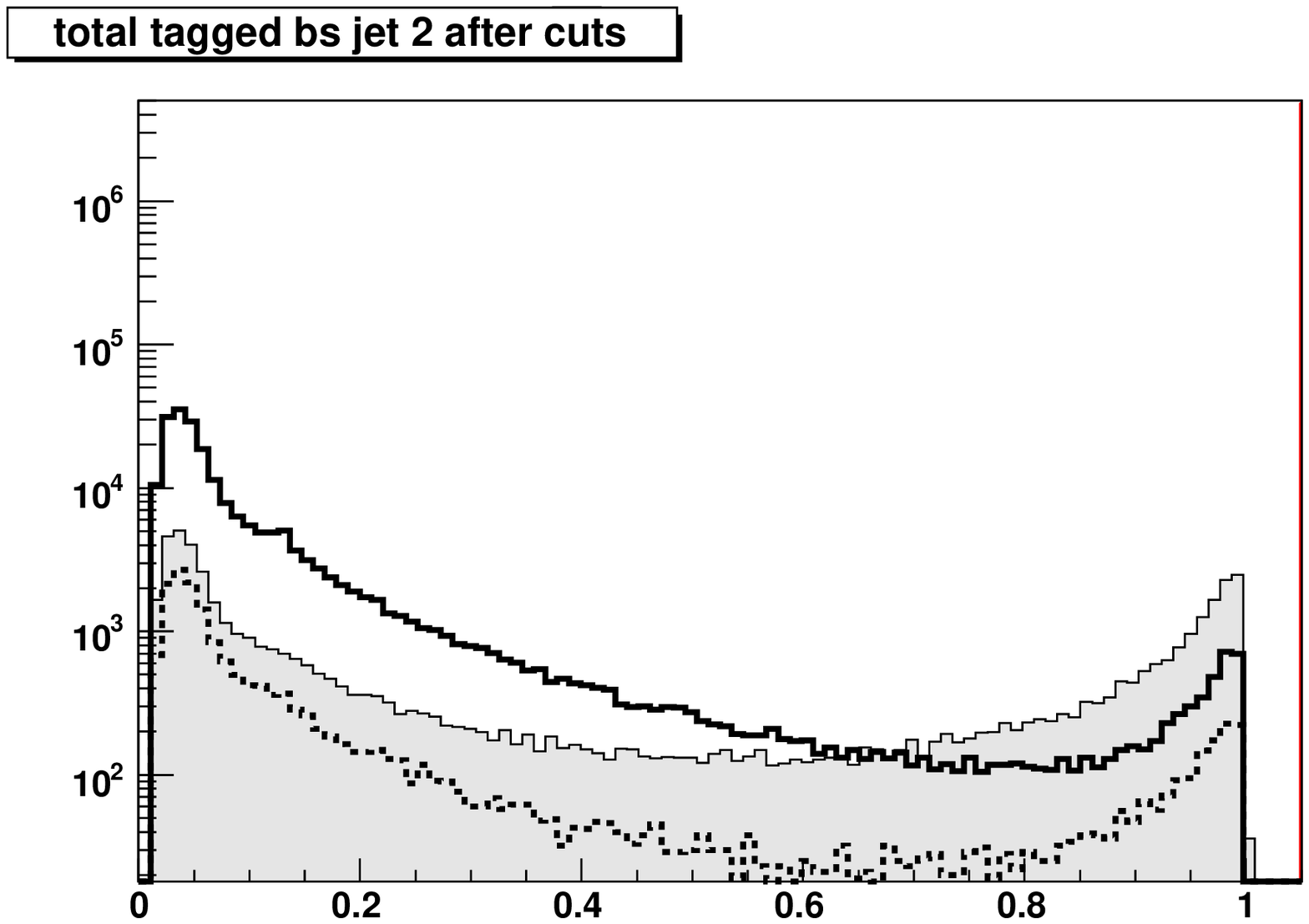}}
\subfloat[]{\includegraphics[scale=0.35]{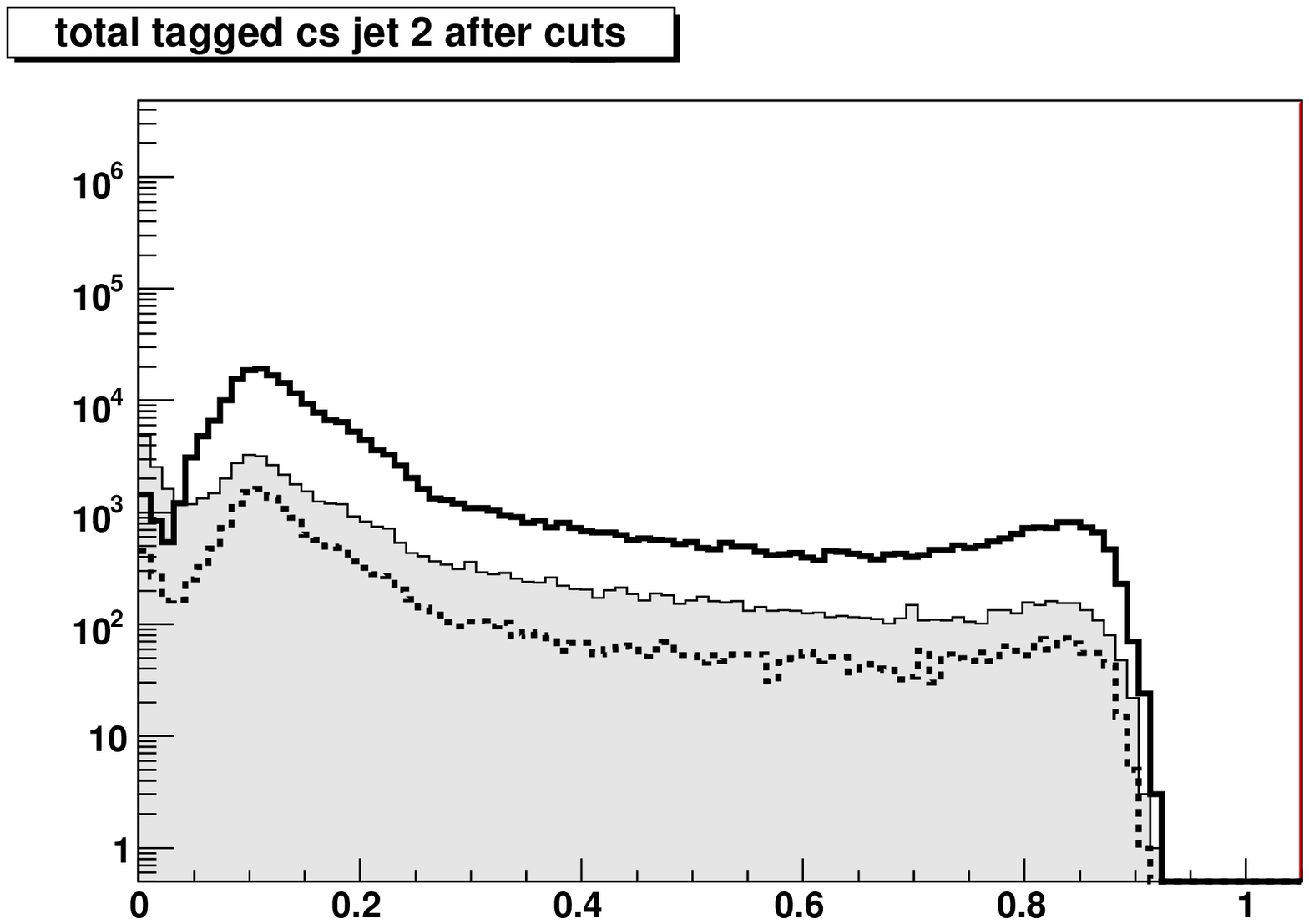}} \\
\subfloat[]{\includegraphics[scale=0.35]{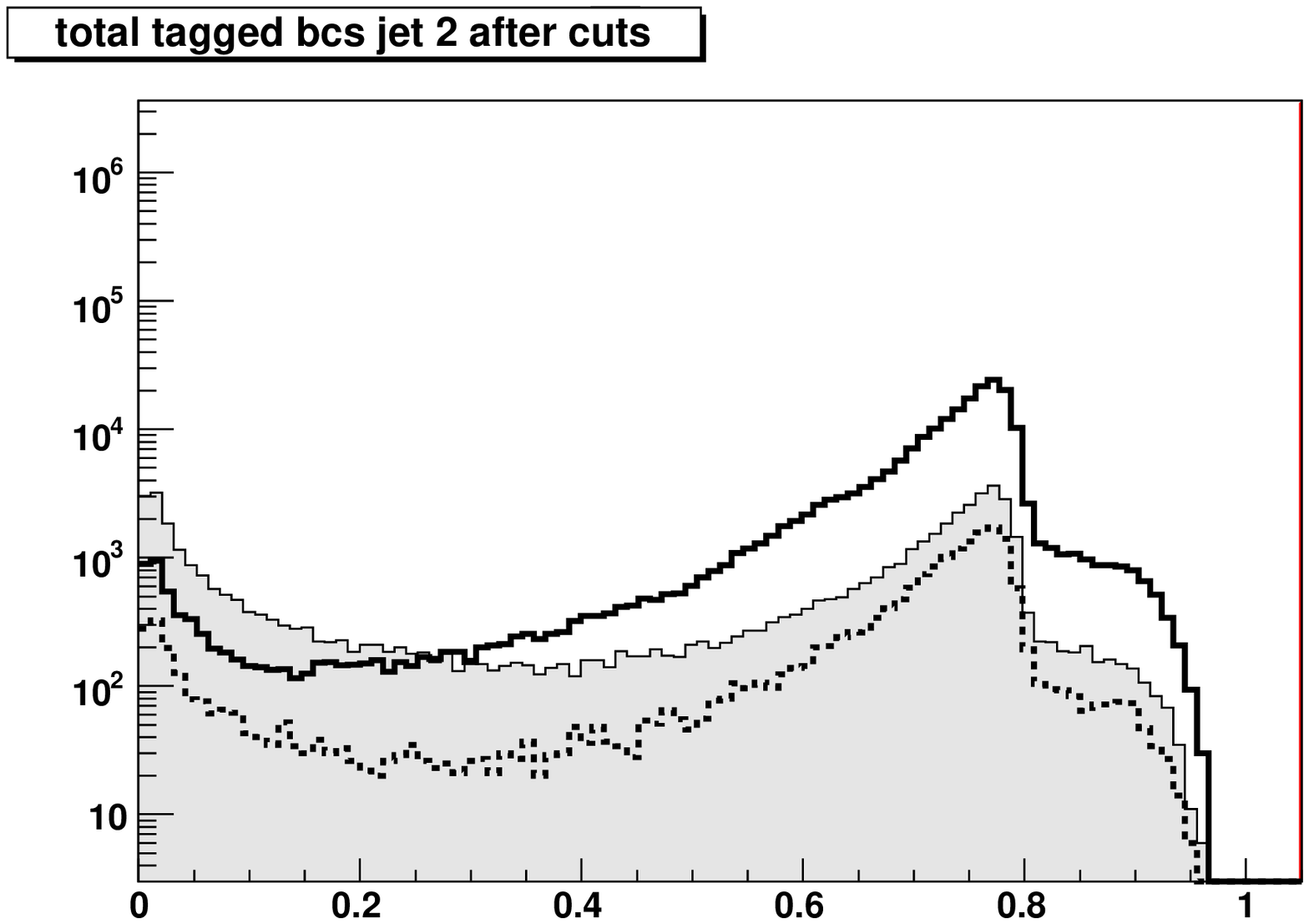}} 
\end{center}
\caption{Distributions of flavour tagging variables for $b\bar{b}$ in hadronic channel: Second jet (a) b-tag; (b) c-tag; (c) c-tag with b background only.}
\label{4bnn2}
\end{figure}

\begin{figure}[htbp]
\begin{center}
\subfloat[]{\includegraphics[scale=0.35]{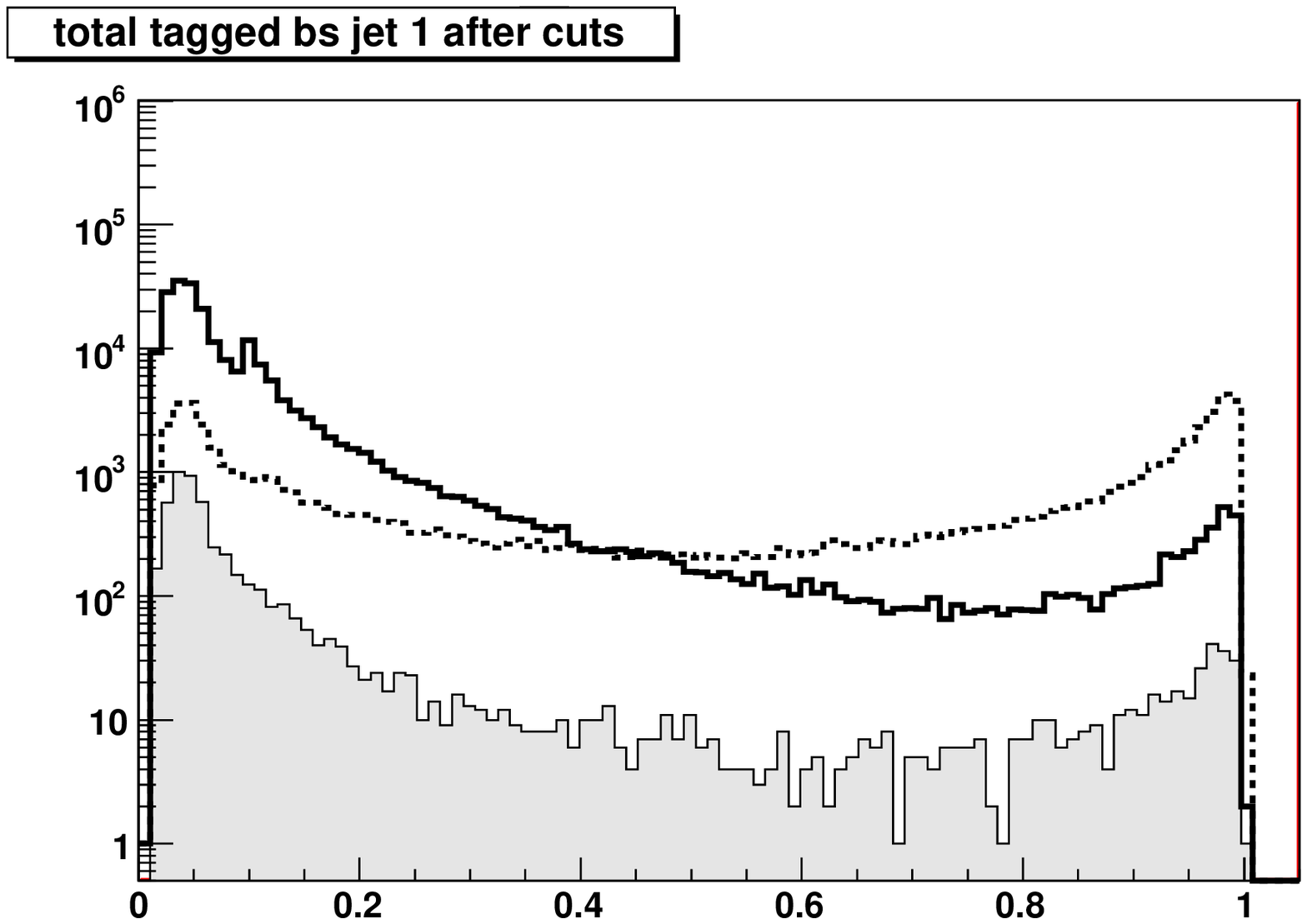}}
\subfloat[]{\includegraphics[scale=0.35]{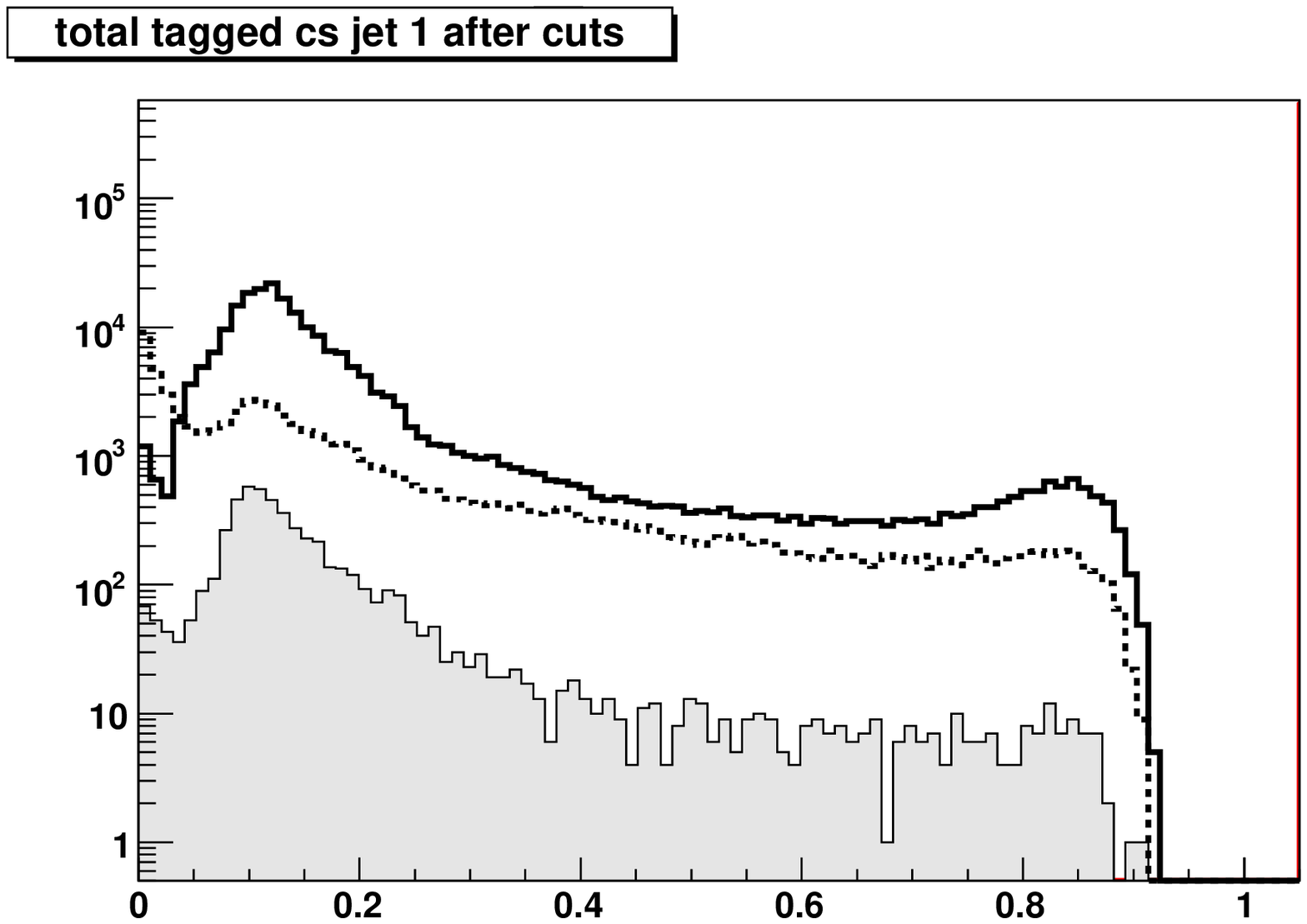}} \\
\subfloat[]{\includegraphics[scale=0.35]{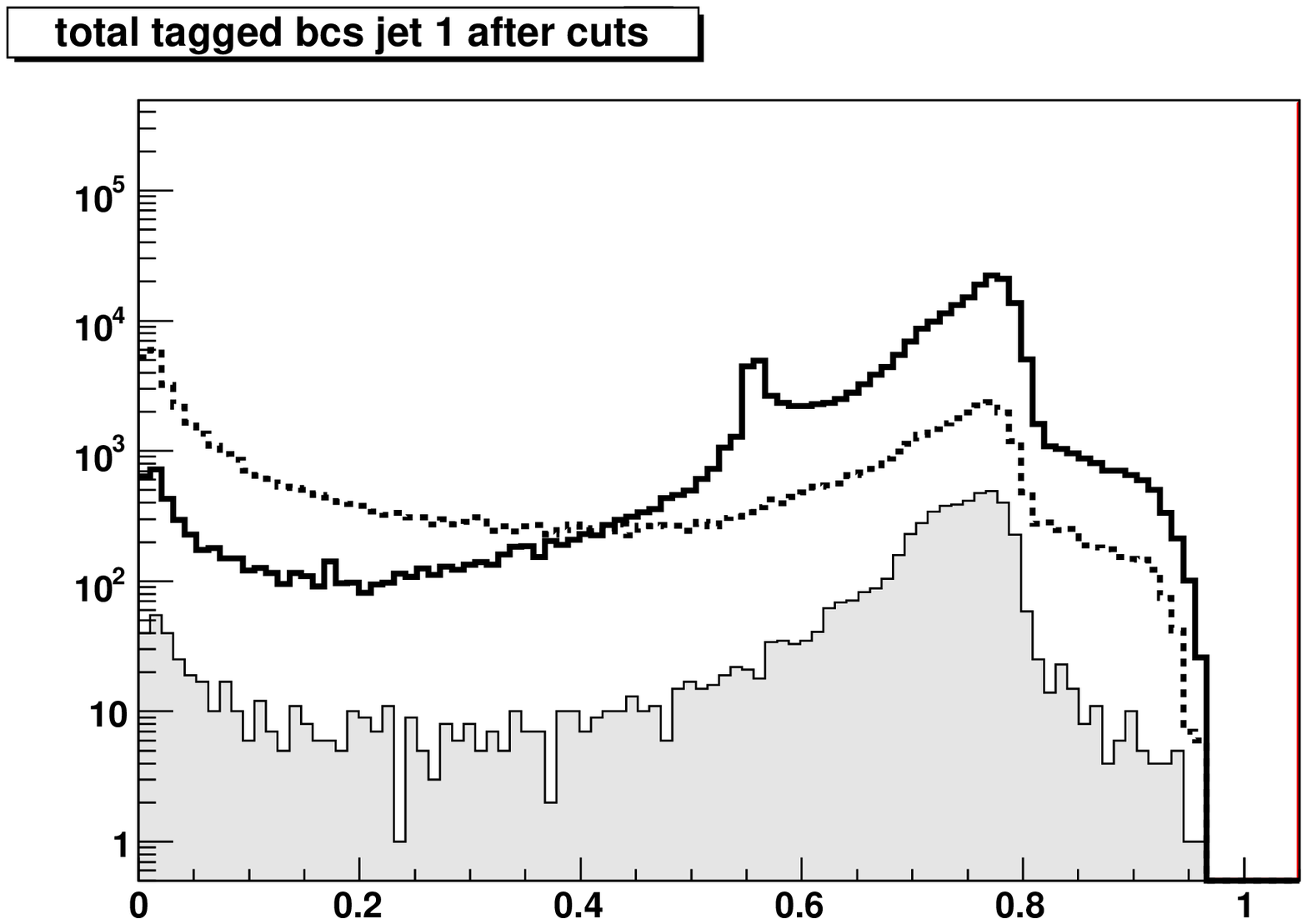}} 
\end{center}
\caption{Distributions of flavour tagging variables for gg in hadronic channel: Leading jet (a) b-tag; (b) c-tag; (c) c-tag with b background only.}
\label{4gnn1}
\end{figure}

\begin{figure}[htbp]
\begin{center}
\subfloat[]{\includegraphics[scale=0.35]{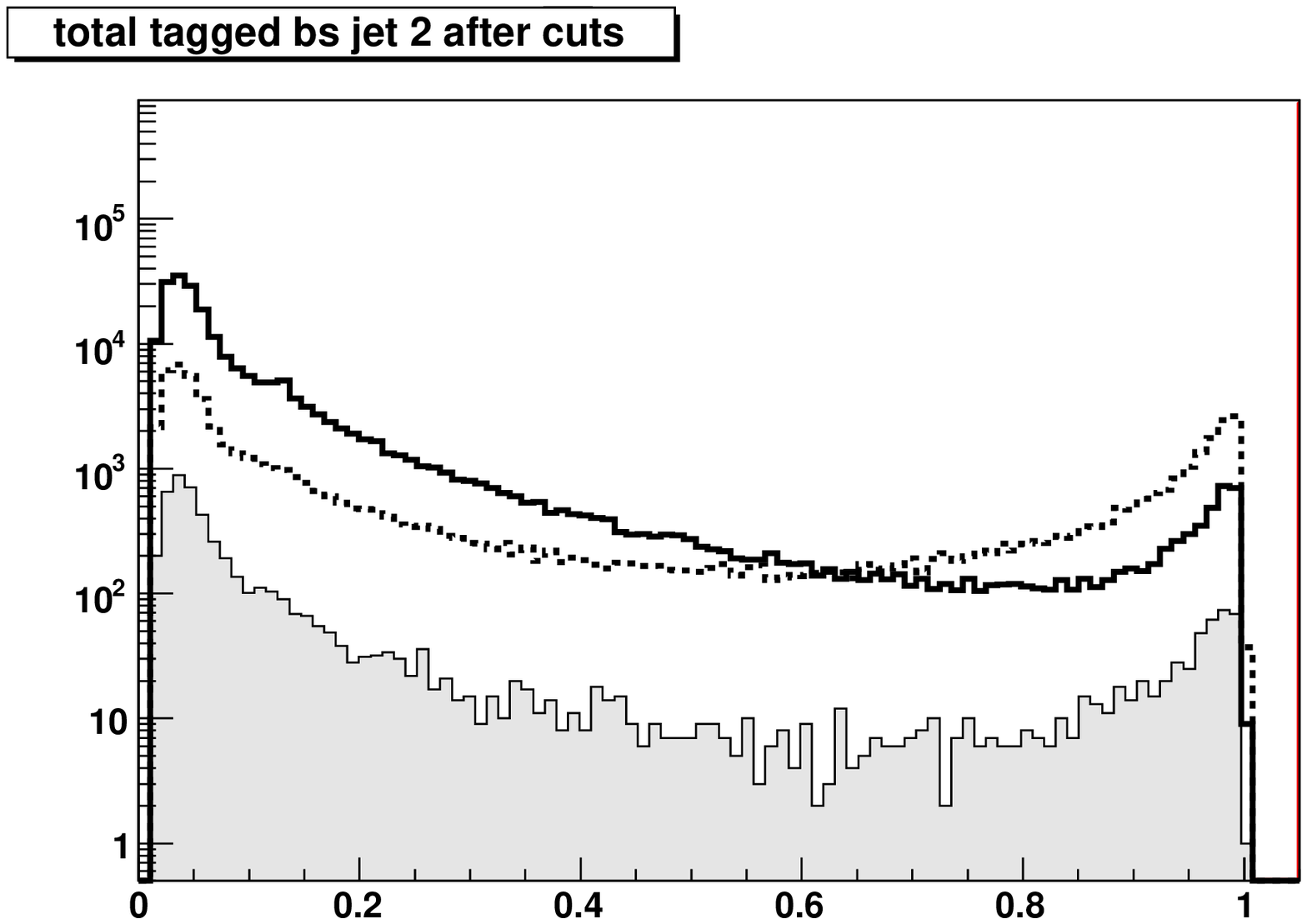}}
\subfloat[]{\includegraphics[scale=0.35]{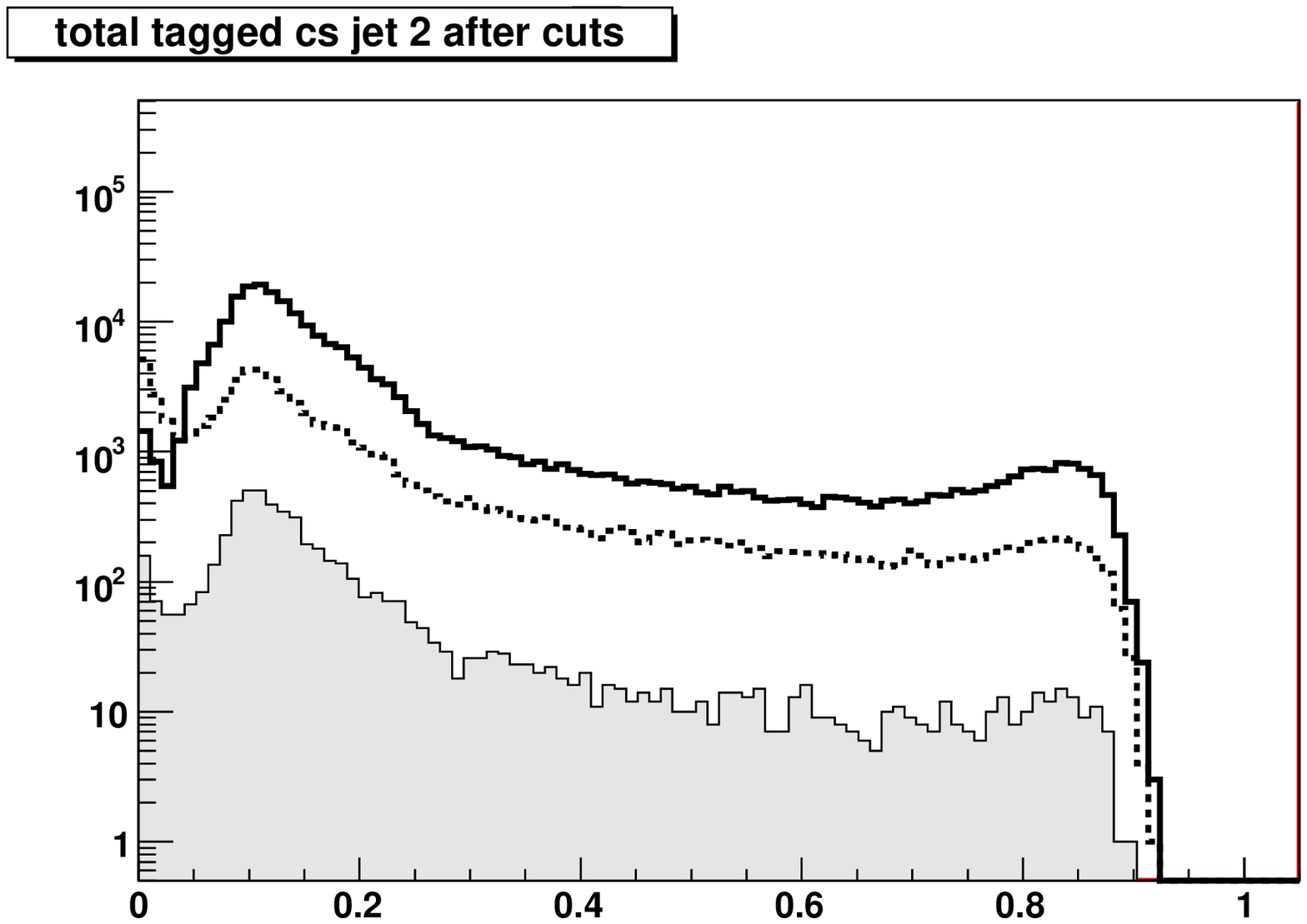}} \\
\subfloat[]{\includegraphics[scale=0.35]{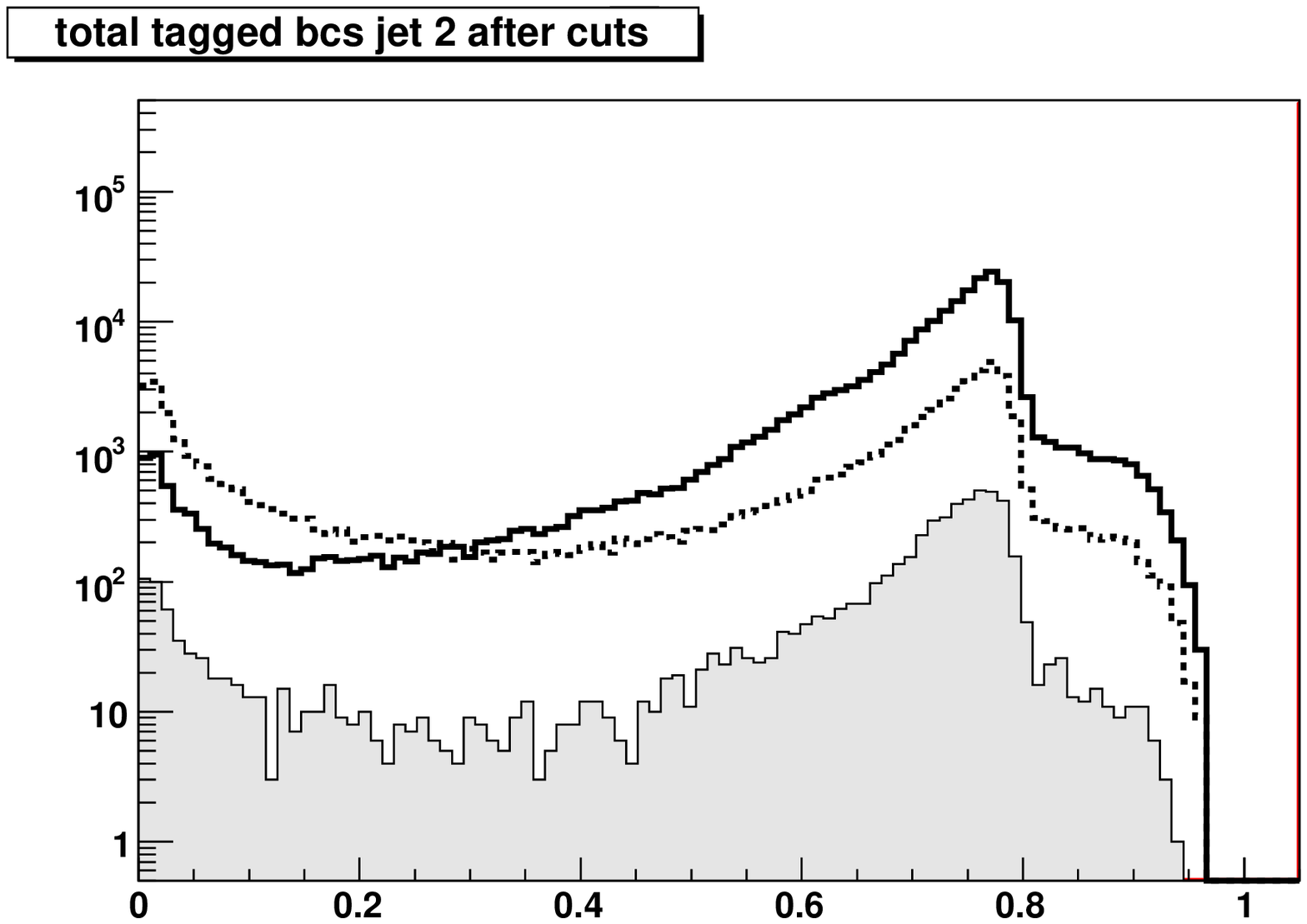}} 
\end{center}
\caption{Distributions of flavour tagging variables for gg in hadronic channel: Second jet (a) b-tag; (b) c-tag; (c) c-tag with b background only.}
\label{4gnn2}
\end{figure}

The main backgrounds at this stage are fully hadronic WW and ZZ pairs, and 2-fermion pairs. As in the neutrino channel, events that survive the pre-selection cuts undergo a neural network analysis to help further discriminate the signal and background. The input variables to the neural network include all pre-selection variables and also the jet flavour tagging outputs produced using the LCFI package. Figures~\ref{4bnn1}, ~\ref{4bnn2}, ~\ref{4gnn1} and ~\ref{4gnn2} show distributions of the three LCFI flavour tags, `b-tag', `c-tag' and  `c-tag with b background only' for the highest (leading) and second highest energy b- and gluon jets.

\subsubsection{Di-jet Mass Resolution}

\begin{figure}[htbp]
\begin{center}
\subfloat[]{\includegraphics[scale=0.35]{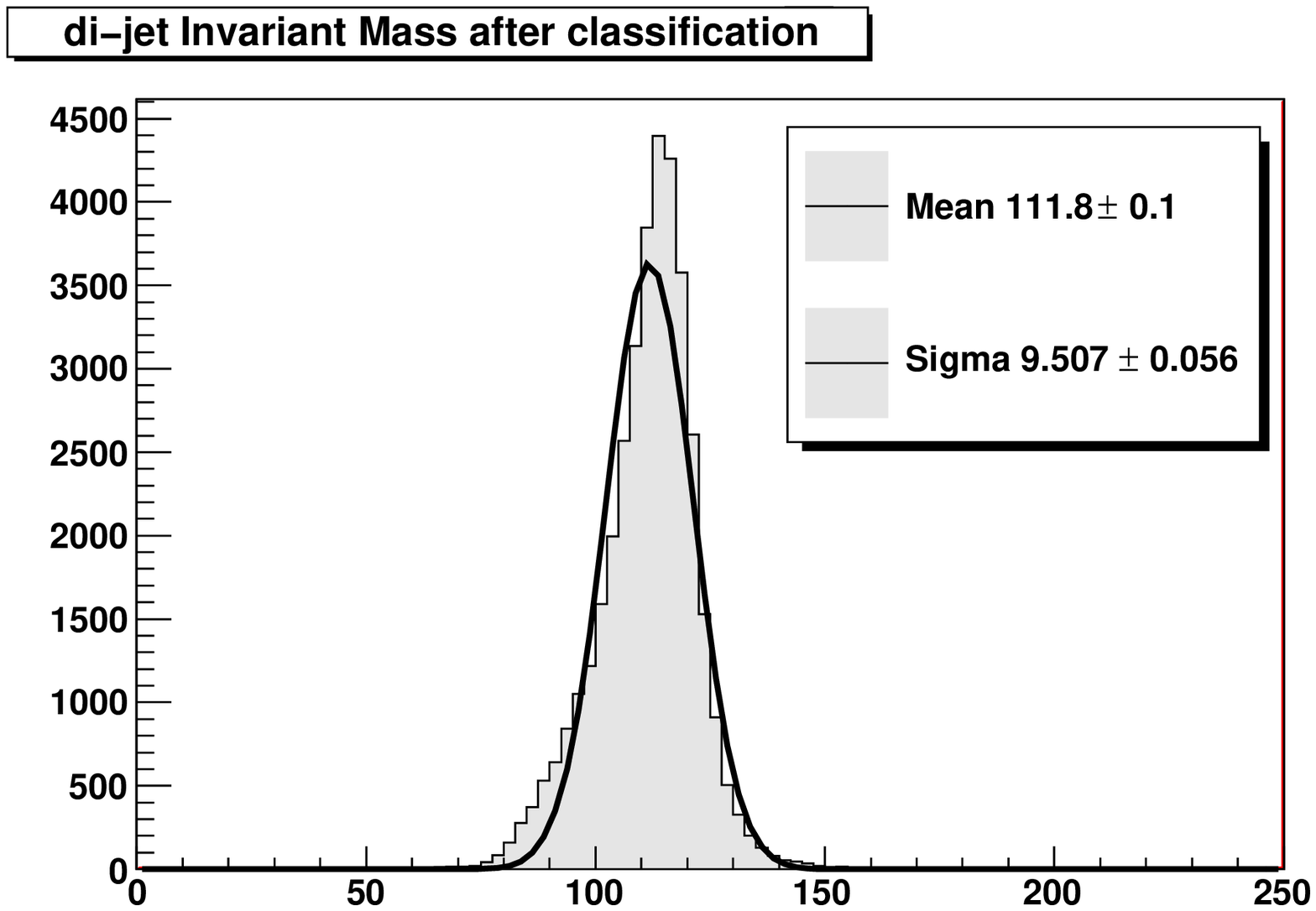}}
\subfloat[]{\includegraphics[scale=0.35]{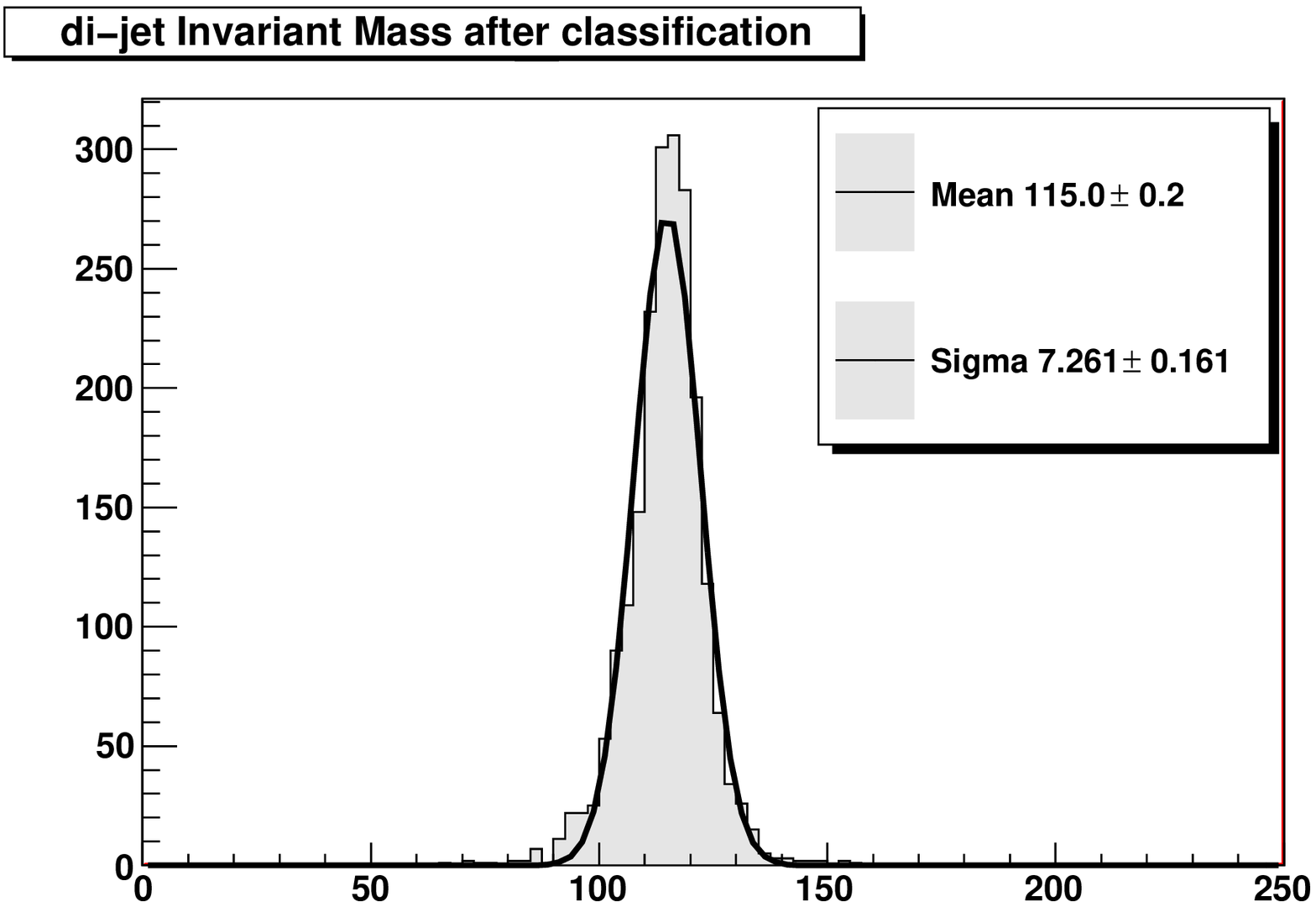}} \\
\subfloat[]{\includegraphics[scale=0.35]{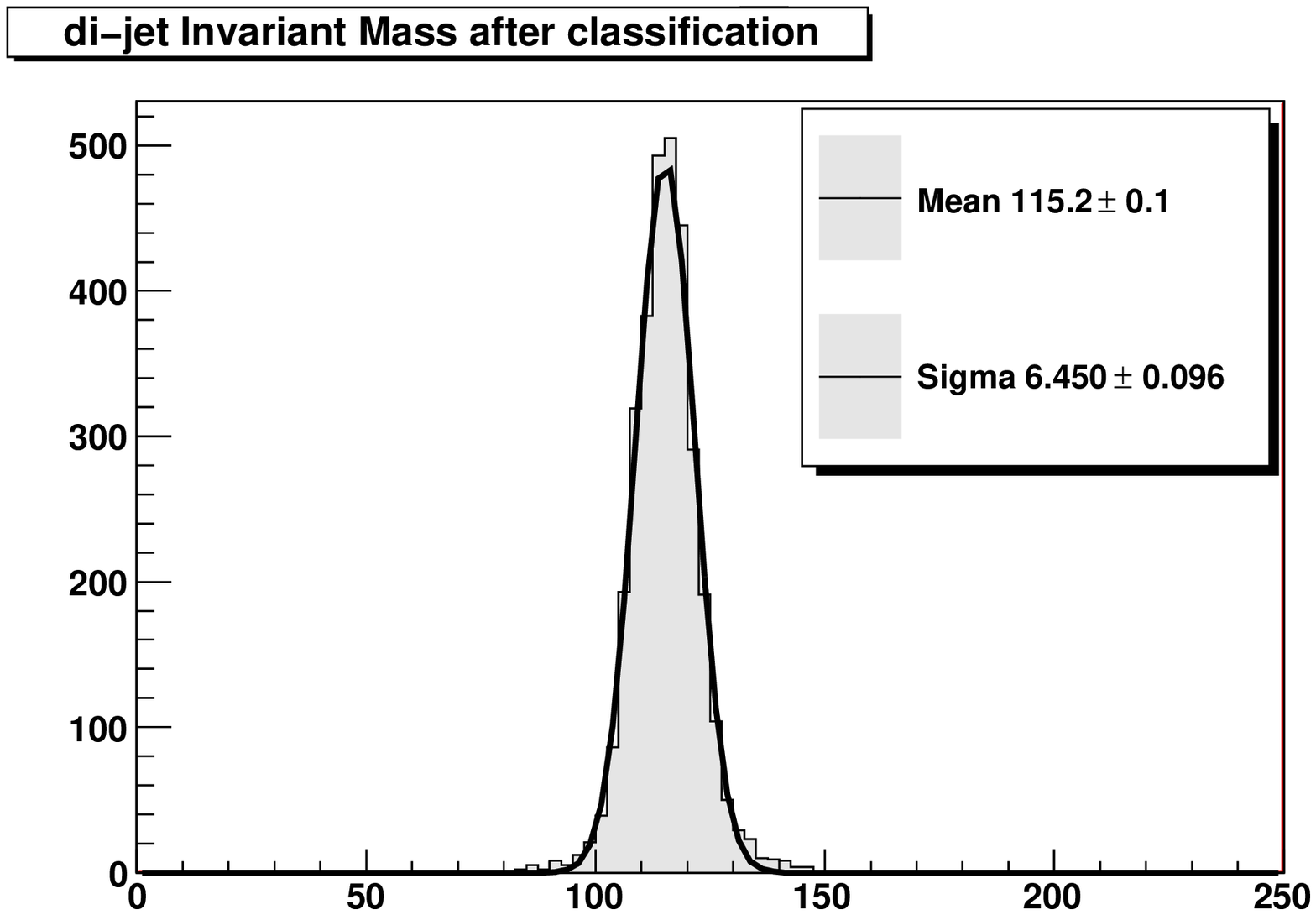}}
\end{center}
\caption{Di-jet invariant mass (GeV) in the neutrino channel: (a) $b\bar{b}$; (b) $c\bar{c}$ and (c) gg.}
\label{numass}
\end{figure}

\begin{figure}[htbp]
\begin{center}
\subfloat[]{\includegraphics[scale=0.35]{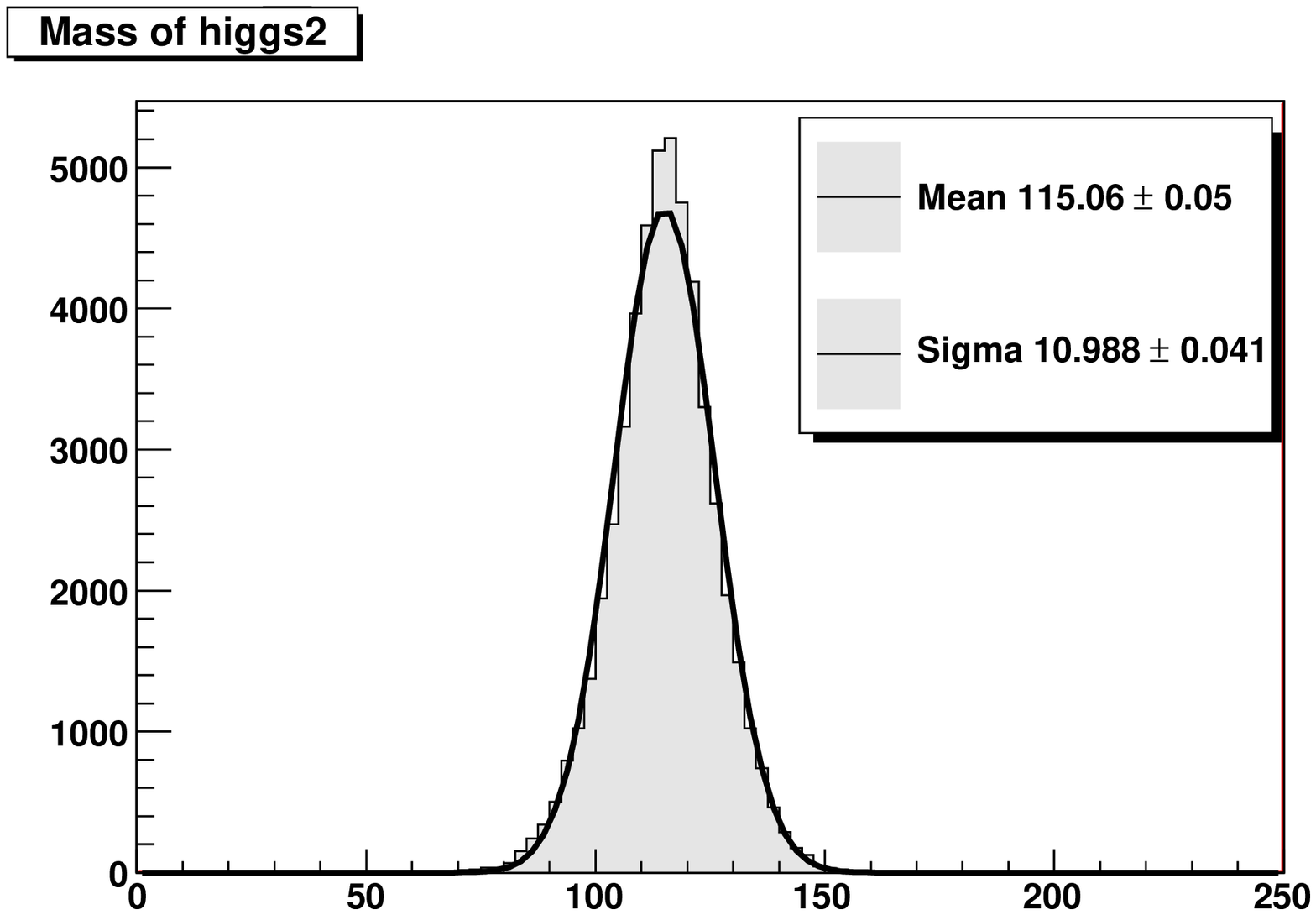}}
\subfloat[]{\includegraphics[scale=0.35]{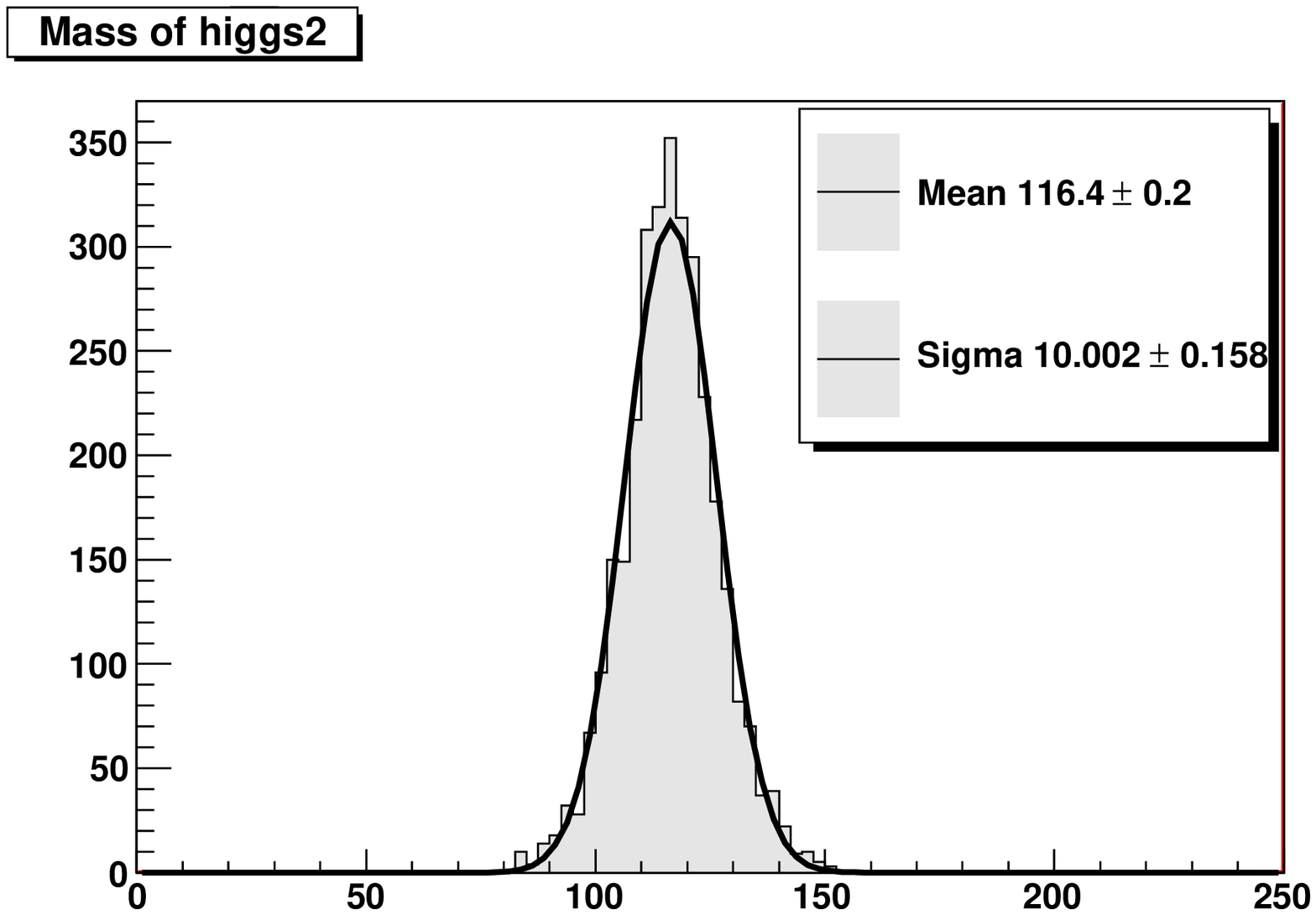}} \\
\subfloat[]{\includegraphics[scale=0.35]{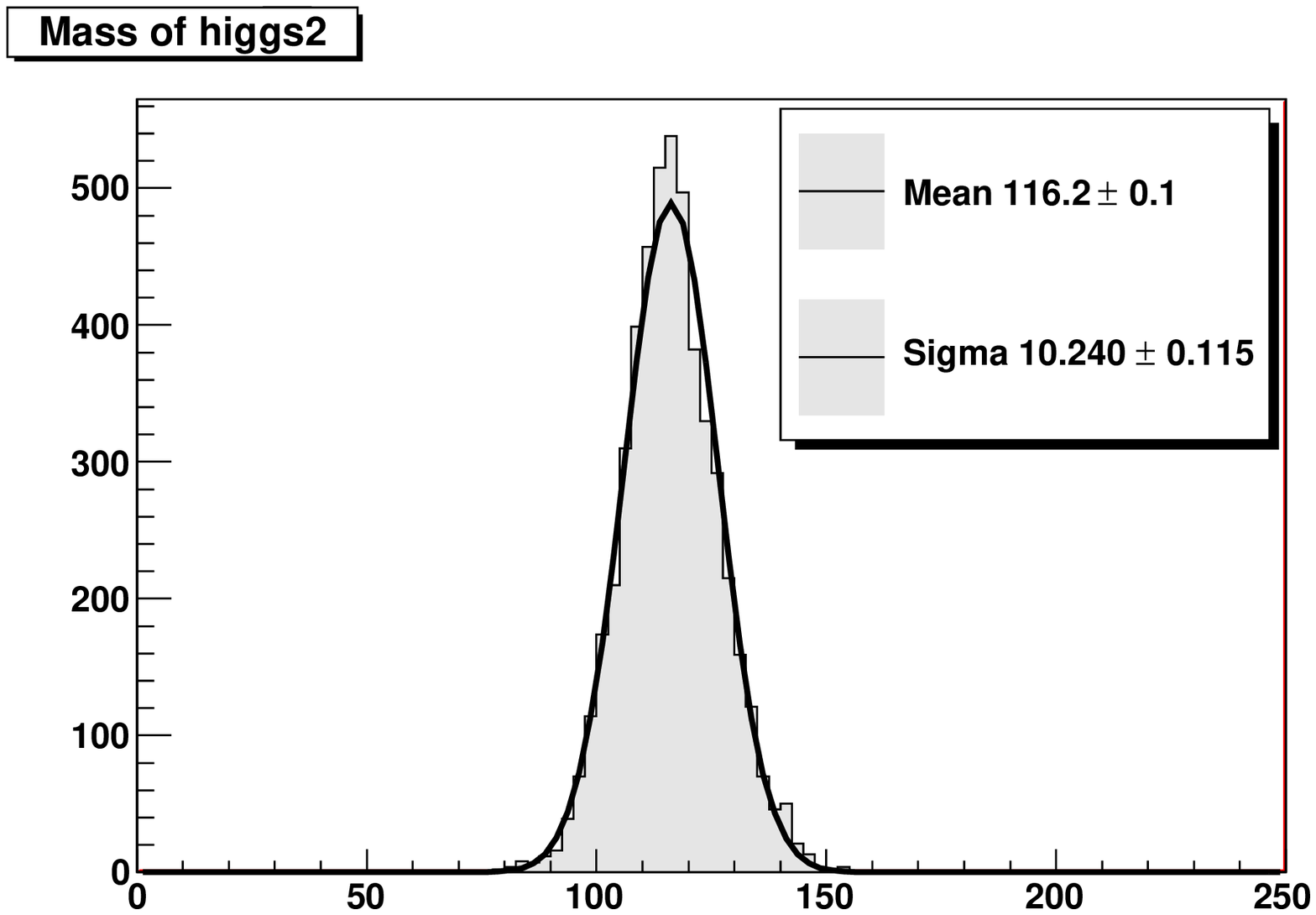}}
\end{center}
\caption{Di-jet invariant mass of the Higgs candidate (GeV) in the hadronic channel before kinematic fitting: (a) $b\bar{b}$; (b) $c\bar{c}$ and (c) gg.}
\label{hadmass}
\end{figure}

\begin{figure}[htbp]
\begin{center}
\subfloat[]{\includegraphics[scale=0.35]{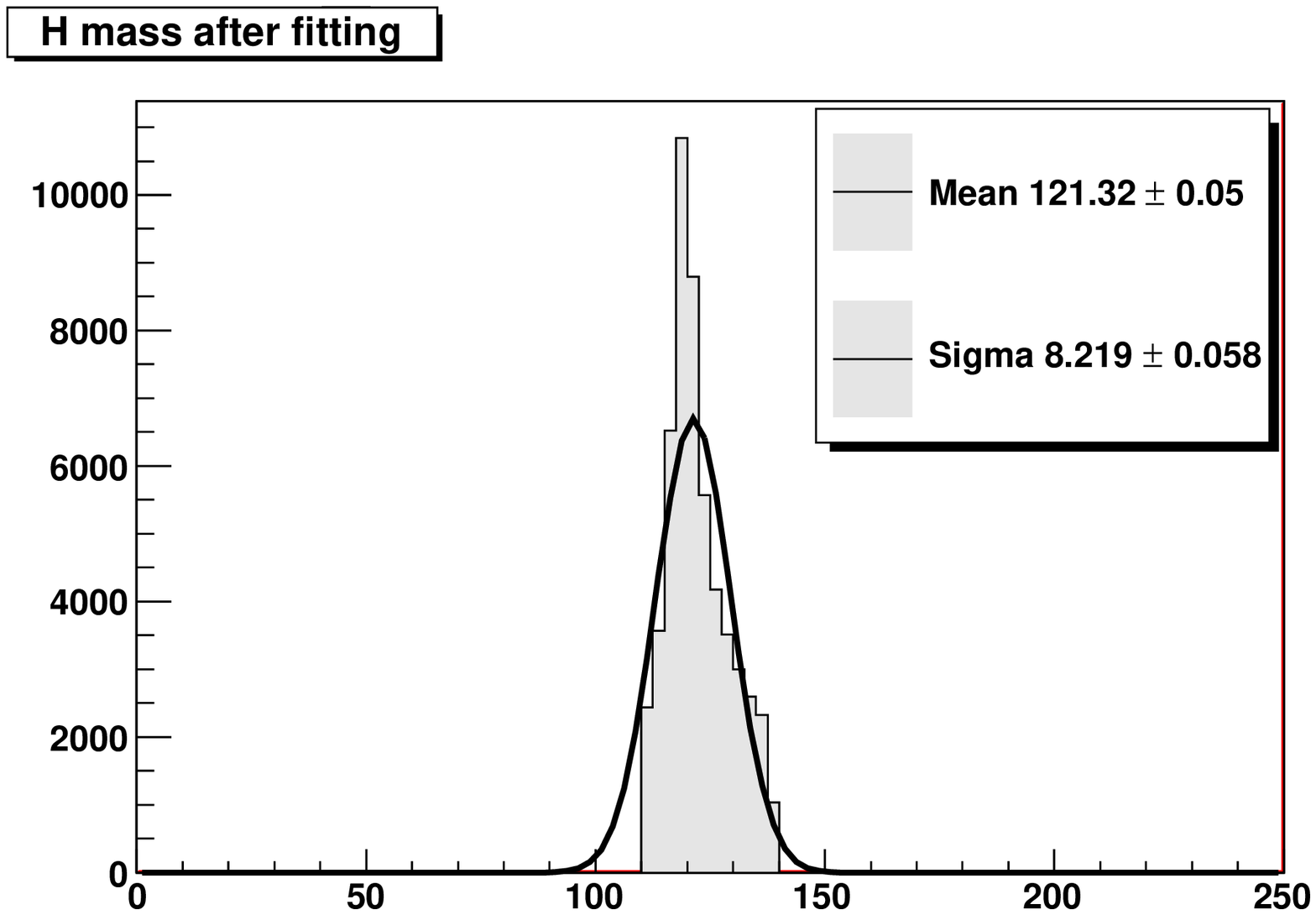}}
\subfloat[]{\includegraphics[scale=0.35]{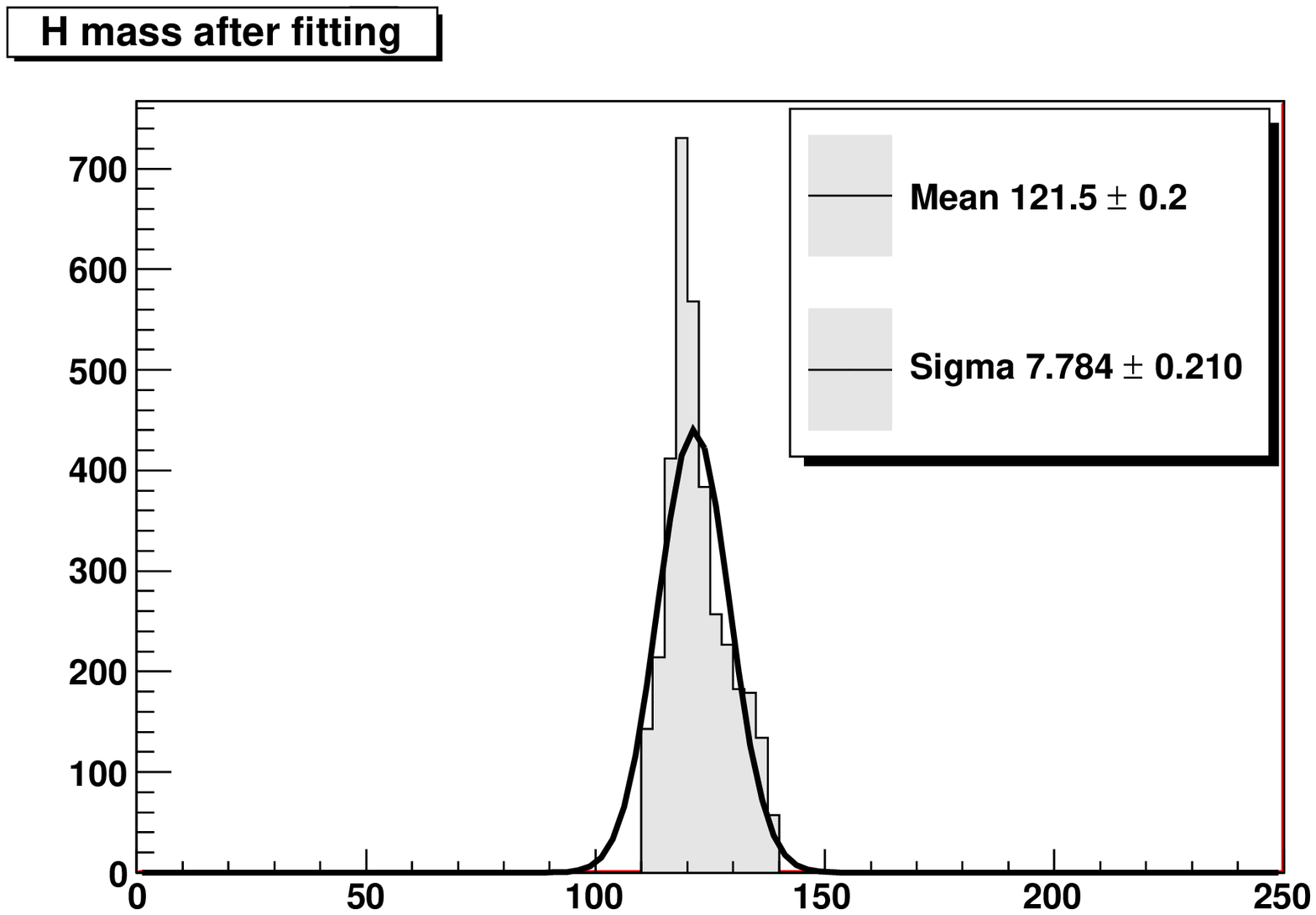}} \\
\subfloat[]{\includegraphics[scale=0.35]{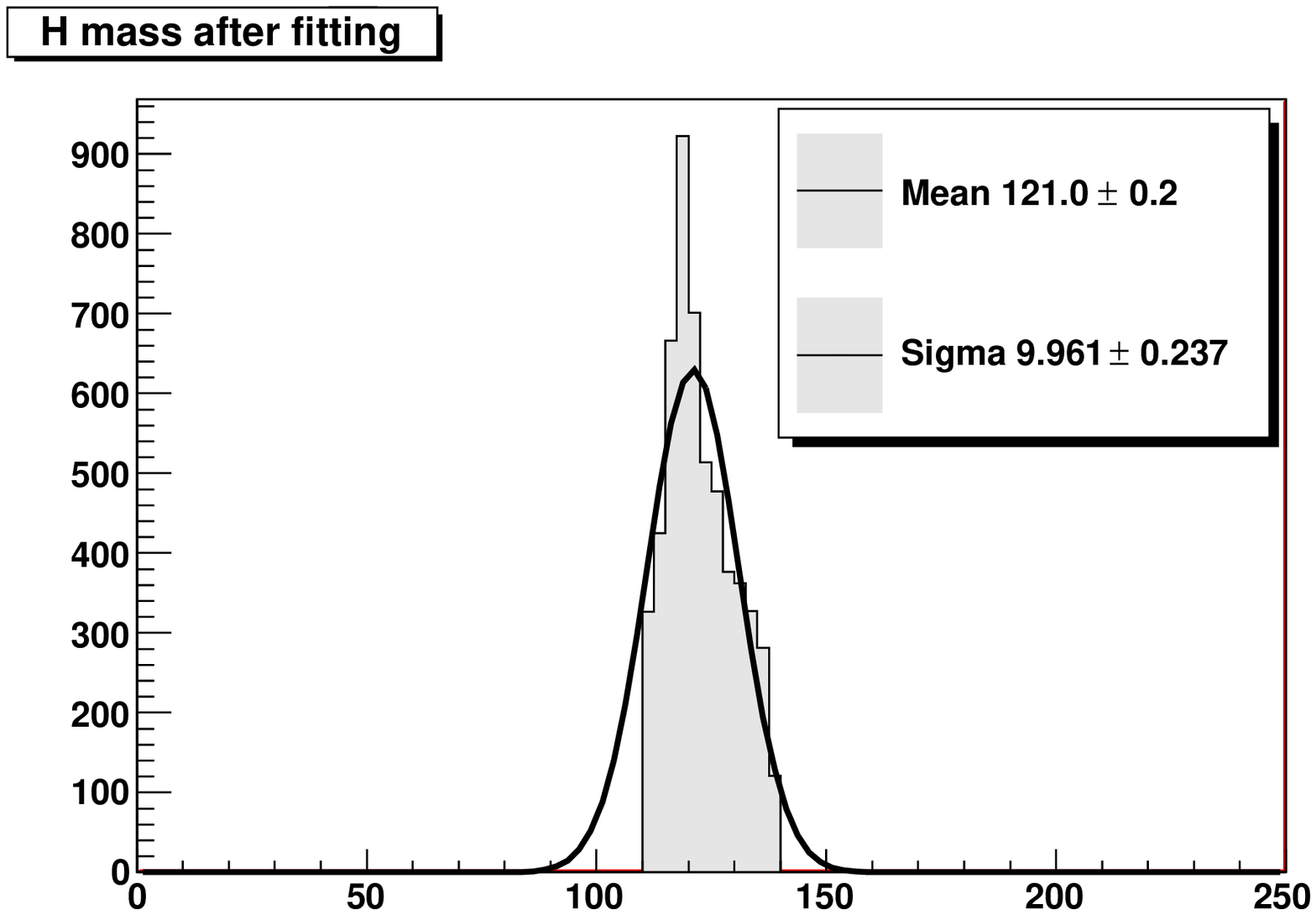}}
\end{center}
\caption{Hadronic channel after kinematic fitting: (a) b di-jet mass; (b) c di-jet mass; (c) gluon di-jet mass.}
\label{fitmass}
\end{figure}

Jet energy measurements are a crucial part of Higgs studies and the measurement of the Higgs mass distribution width is important. Figures~\ref{numass} and 
~\ref{hadmass} show the reconstructed visible mass distribution of Higgs decays to $b\bar{b}$, $c\bar{c}$ and gg in the neutrino and hadronic channels 
respectively. The histograms are fit with a single gaussian which is not always adequate but nevertheless allows qualitative analysis. In the neutrino channel the $b\bar{b}$ and $c\bar{c}$ systems have broader mass widths than the gluons because of extra neutrinos coming from b- abd c-hadron semi-leptonic decays. The gluon system has a smaller width in the neutrino channel compared to the hadronic channel where the width of the gluon system is broader due to combinatorics. This information was not explicitly used in the analysis. Figure~\ref{fitmass} shows the $b\bar{b}$, $c\bar{c}$ and gg Higgs mass distribution in the hadronic channel after kinematic fitting.  

\subsection{Determination of Branching Ratios}

The branching ratio of the Higgs boson decay to quarks and gluons was calculated using events that passed the final neural network 
selection. The calculation was done by normalising the signal cross section to the inclusive Higgs cross section, $\sigma_{ZH}$ = 209$\pm$9.8 fb, 
as determined in an independent recoil mass analysis performed for the SiD Letter of Intent~\cite{sidloi}. The branching ratio is then given by
\begin{equation}
BR(H \rightarrow f\bar{f}) = \frac{\sigma_{Hff}}{\sigma_{ZH}}
\end{equation}
where $f$ represents the daughter decay products from the Higgs.
The relative accuracy of the the branching ratio takes into account both the relative signal cross section uncertainty and
the relative Higgs-strahlung uncertainty given as
\begin{equation}
\frac{\Delta BR}{BR} = \sqrt{\left(\frac{\Delta\sigma_{Hff}}{\sigma_{Hff}}\right)^2 + \left(\frac{\Delta\sigma_{ZH}}{\sigma_{ZH}}\right)^2}
\end{equation}
with the relative signal cross section uncertainty calculated by
\begin{equation}
\frac{\Delta\sigma_{Hff}}{\sigma_{Hff}} = \frac{\sqrt{signal + background}}{signal}
\end{equation}
and the cross-section is calculated as follows
\begin{equation}
\sigma_{Hff} = \frac{N}{\varepsilon_{Hff} L}
\end{equation}
where $N$ is the number of signal events after all selections, $\varepsilon$ is the efficiency of signal selection and $L$ is the total integrated luminosity.

The uncertainty in the efficiency is considered negligible, relying on simulations to determine it with sufficient precision. The 
systematic effects or contributions of the luminosity uncertainty were not considered in this analysis.

The weighted average of the signal cross section and its uncertainty are calculated using cross section and relative uncertainty values obtained from the neutrino and hadronic channels. The weighted average cross section is given by
\begin{equation}
\sigma_{average} = \frac{x(\delta y)^2 + y(\delta x)^2}{(\delta x)^2 + (\delta y)^2},
\end{equation}
where $x$ and $y$ are the cross sections in the neutrino and hadronics channels respectively, and $\delta x$ and $\delta y$ are the cross section 
uncertainties in the neutrino and hadronic channels respectively assuming that the two channels are statistically independent. The uncertainty of the average cross section is then calculated as
\begin{equation}
\delta z = \frac{\delta x*\delta y}{\sqrt{(\delta x)^2 + (\delta y)^2}},
\end{equation}
where $\delta z$ is the uncertainty of the weighted average cross section.

\begin{figure}[htbp]
\begin{center}
\subfloat[]{\includegraphics[scale=0.25]{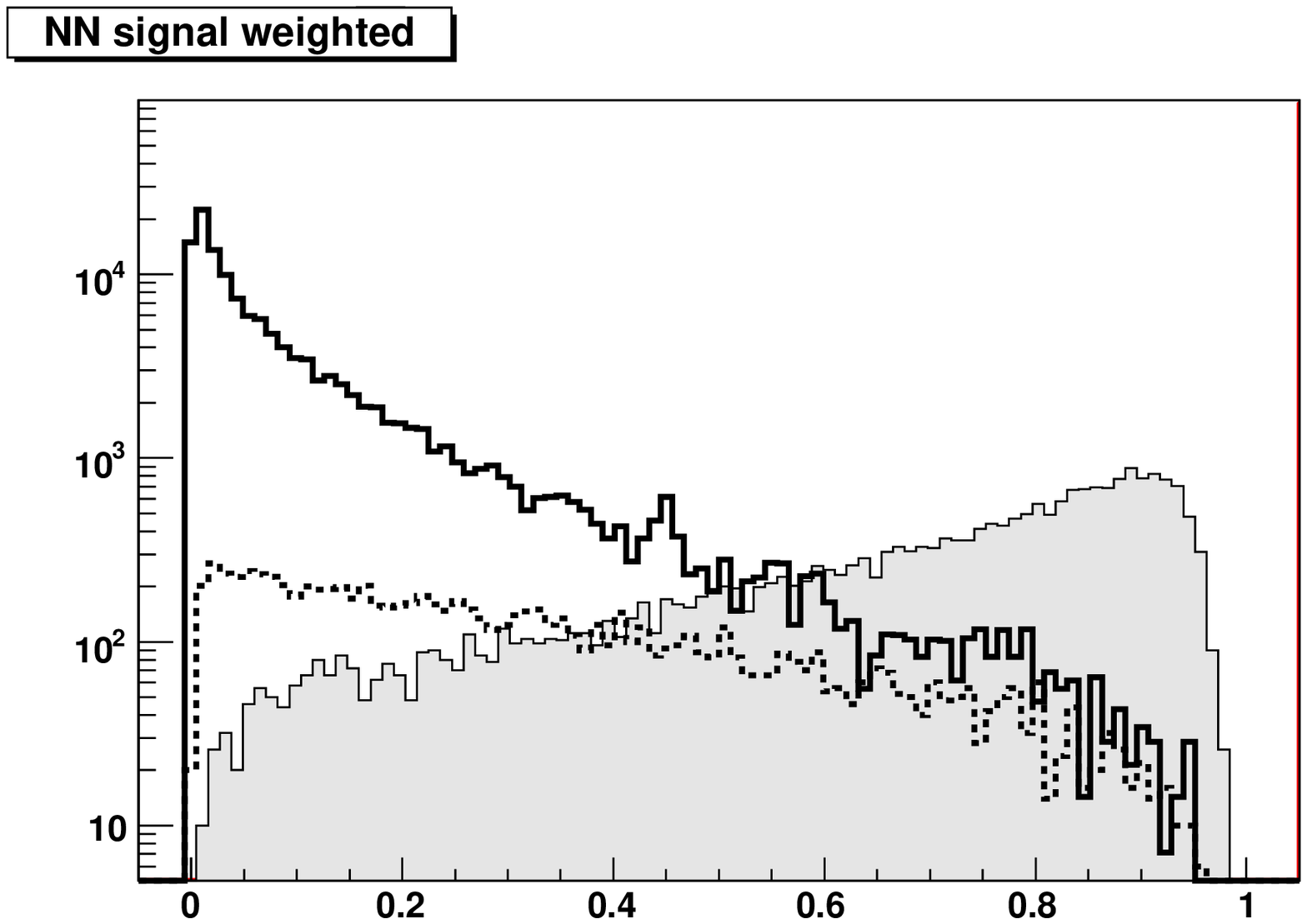}}
\subfloat[]{\includegraphics[scale=0.25]{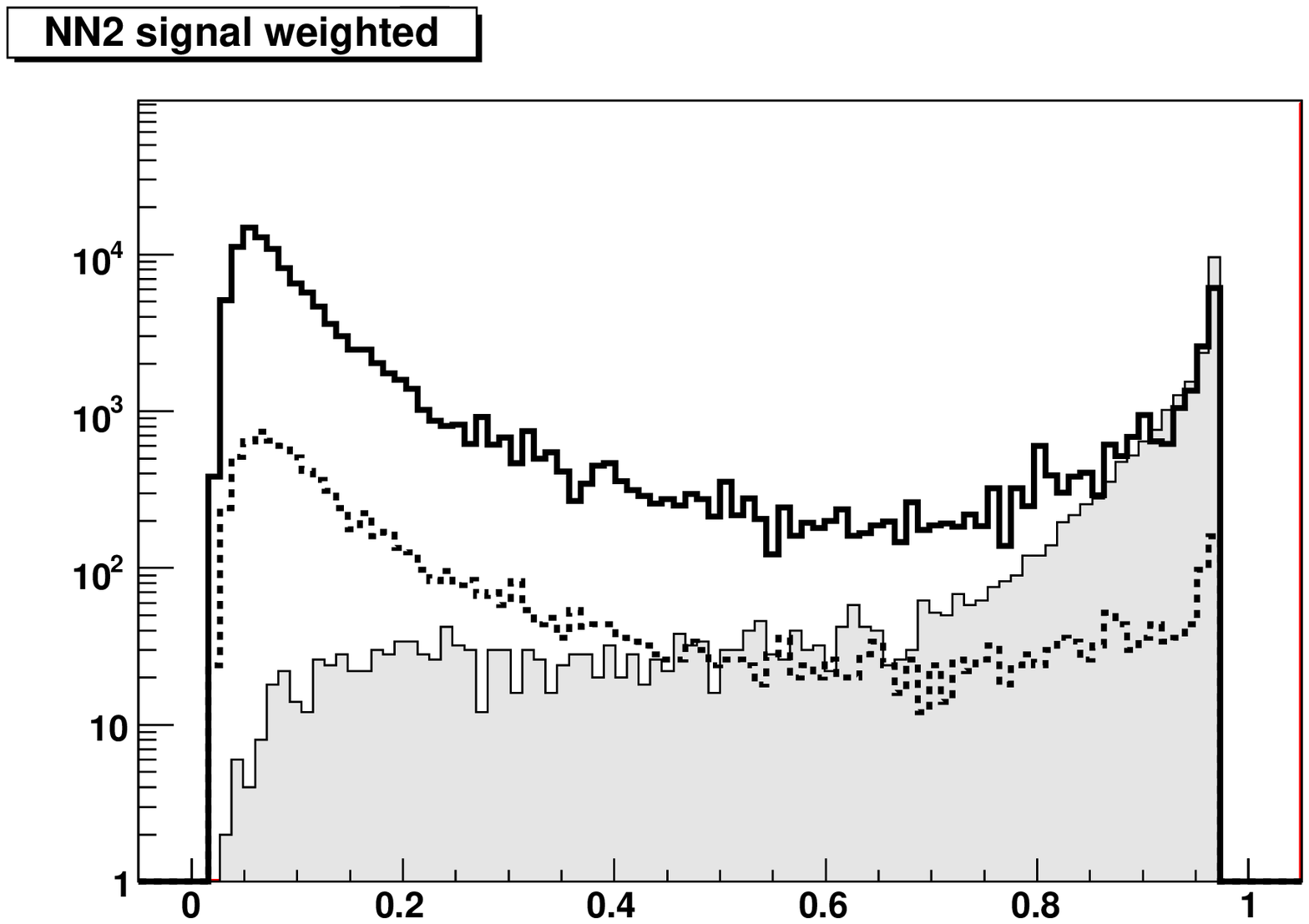}} \\
\subfloat[]{\includegraphics[scale=0.25]{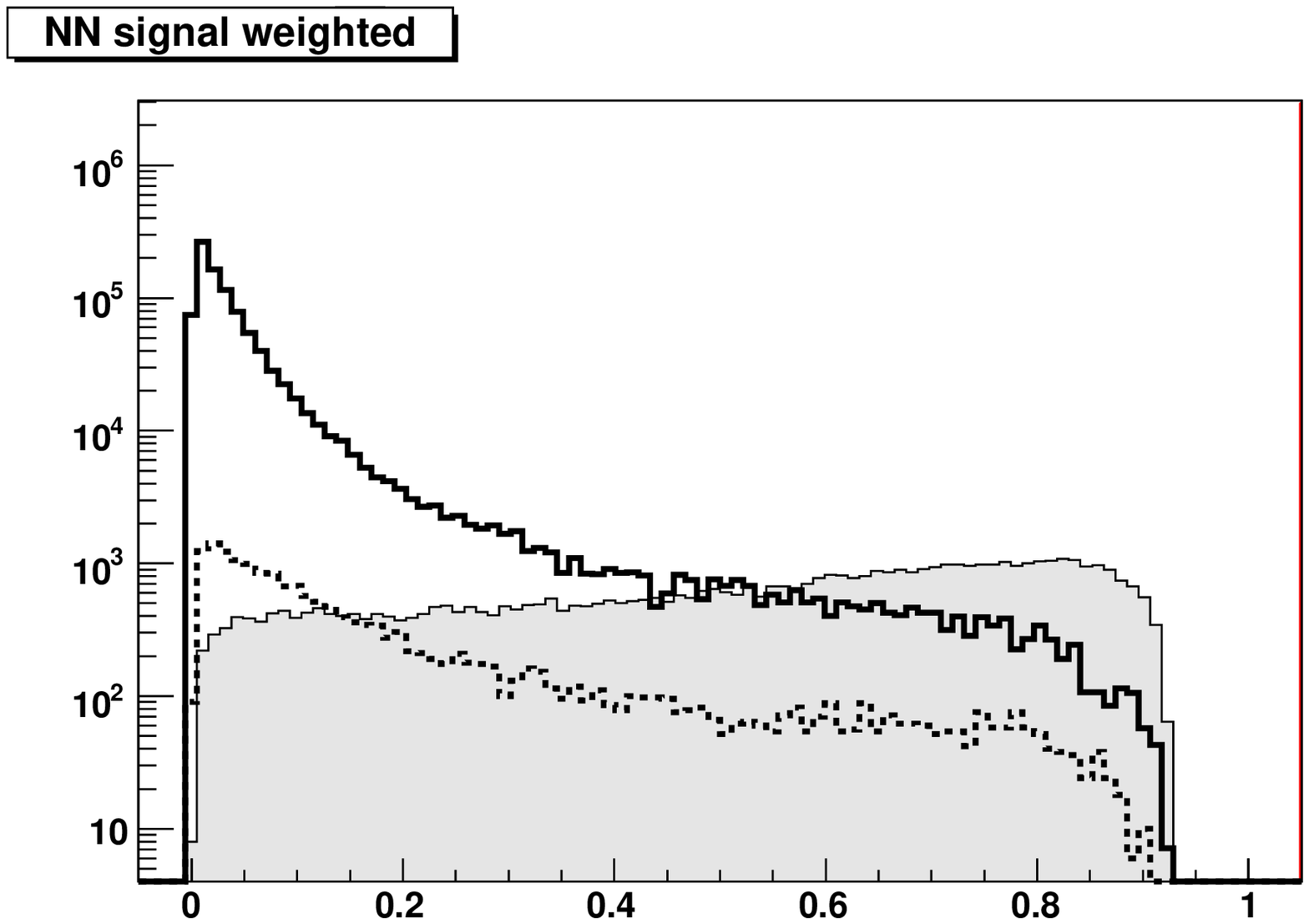}}
\subfloat[]{\includegraphics[scale=0.25]{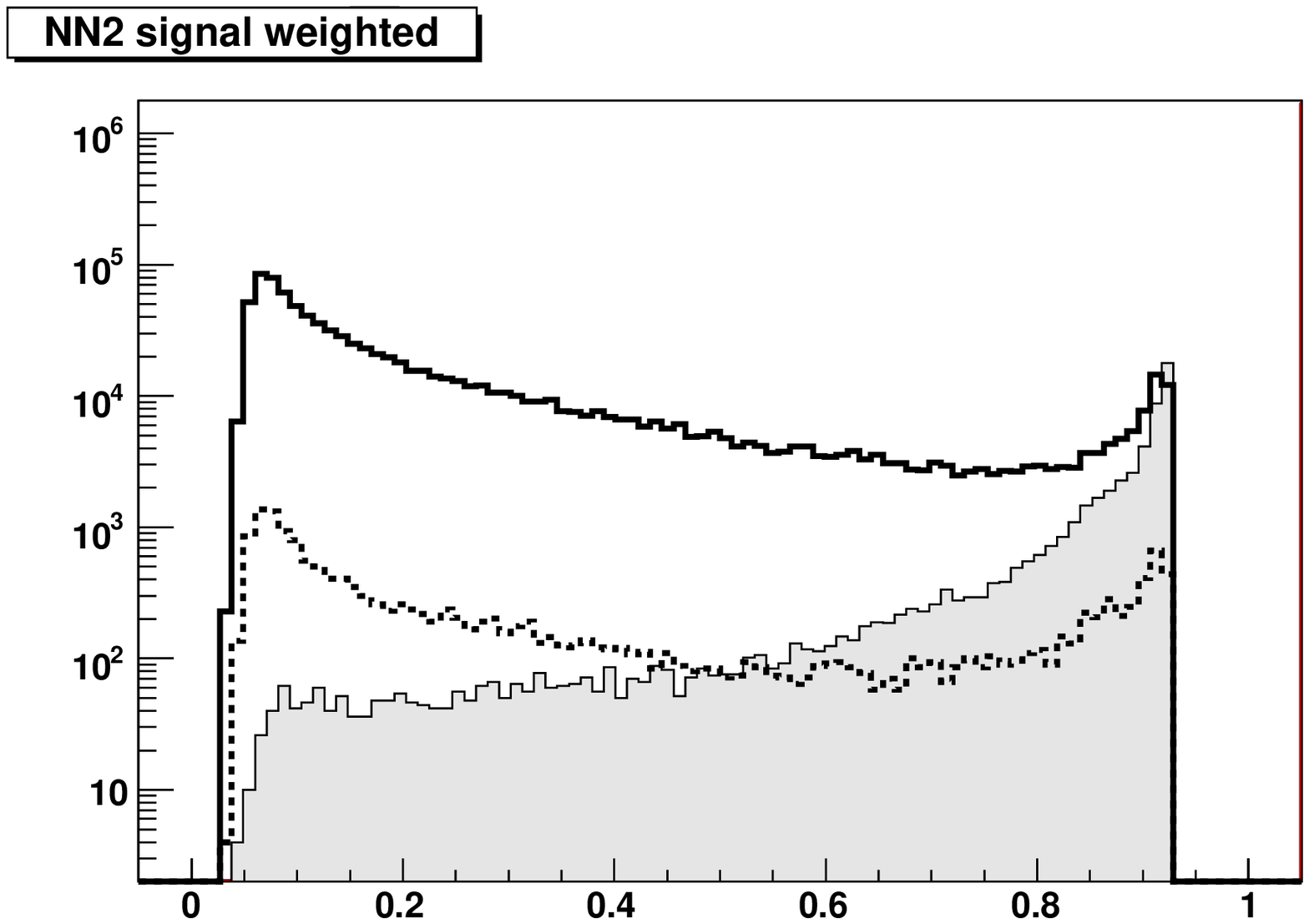}}
\end{center}
\caption{$b\bar{b}$: Neutrino channel (a) and (b), and Hadronic channel (c) and (d). First NN corresponds to (a) and (c), and Second NN corresponds to (b) and (d). Solid curves are SM background, dashed curves are inclusive Higgs sample (with removed signal) and filled histograms are the signals.}
\label{NNb}
\end{figure}

\begin{figure}[htbp]
\begin{center}
\subfloat[]{\includegraphics[scale=0.25]{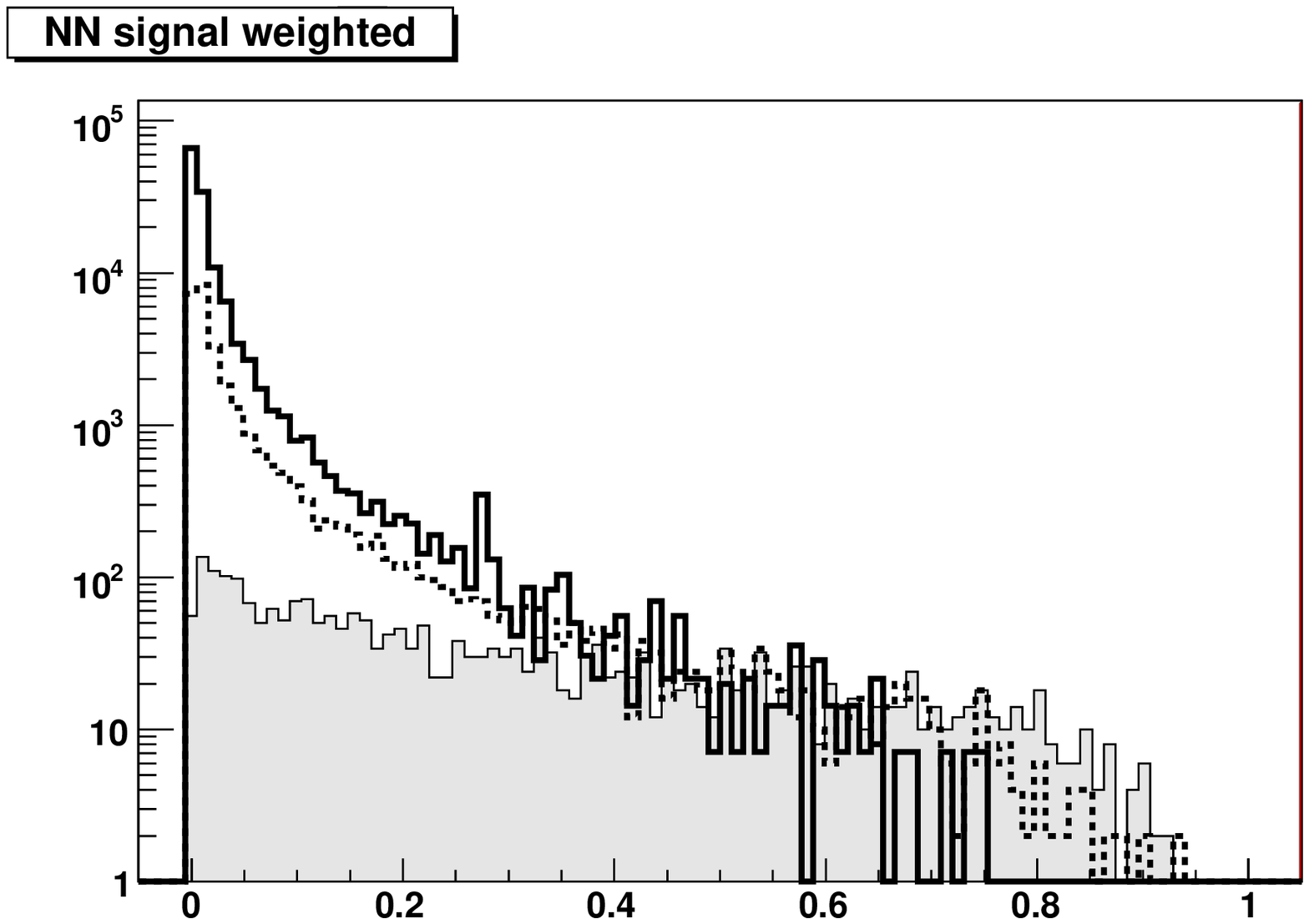}}
\subfloat[]{\includegraphics[scale=0.25]{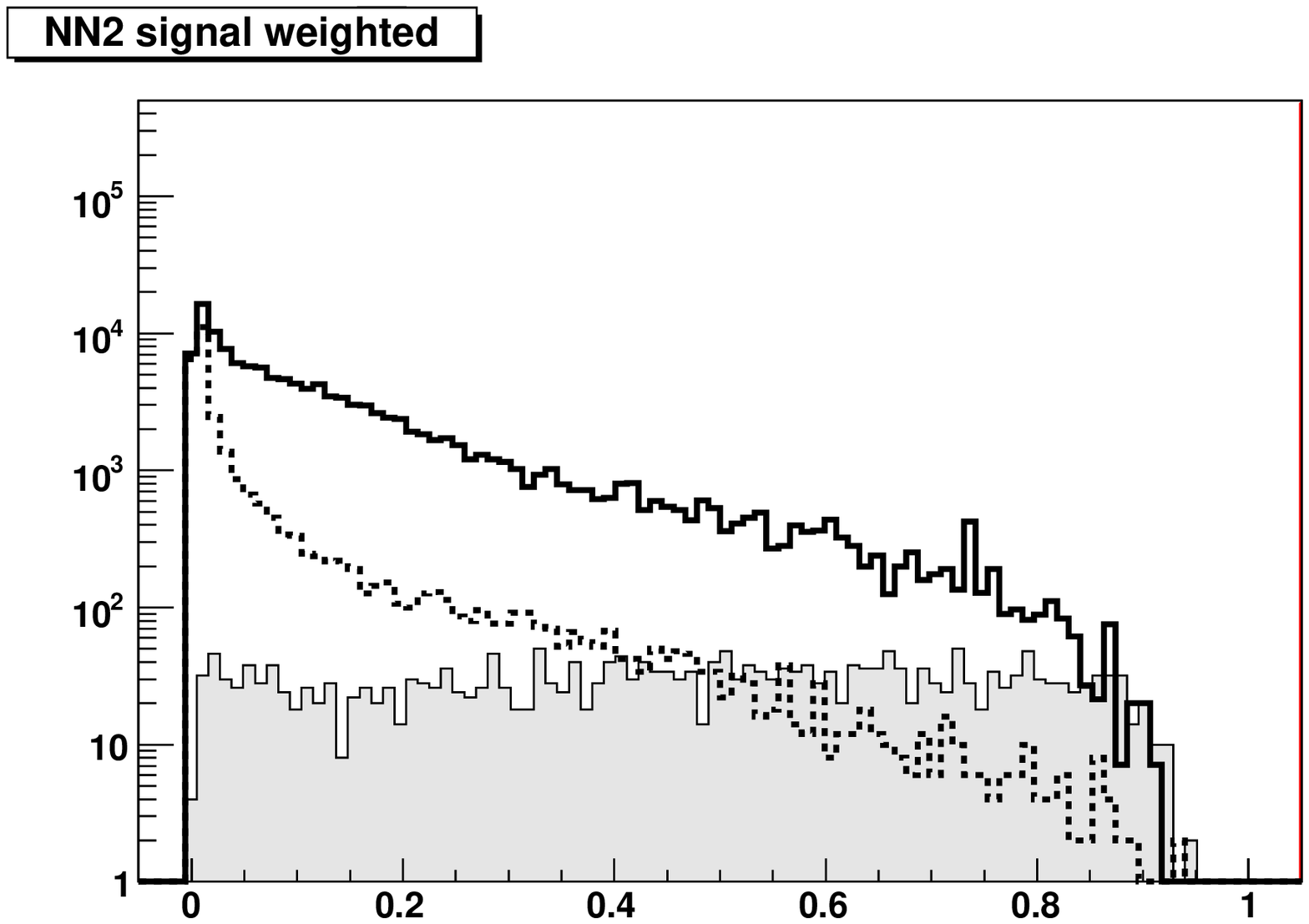}} \\
\subfloat[]{\includegraphics[scale=0.25]{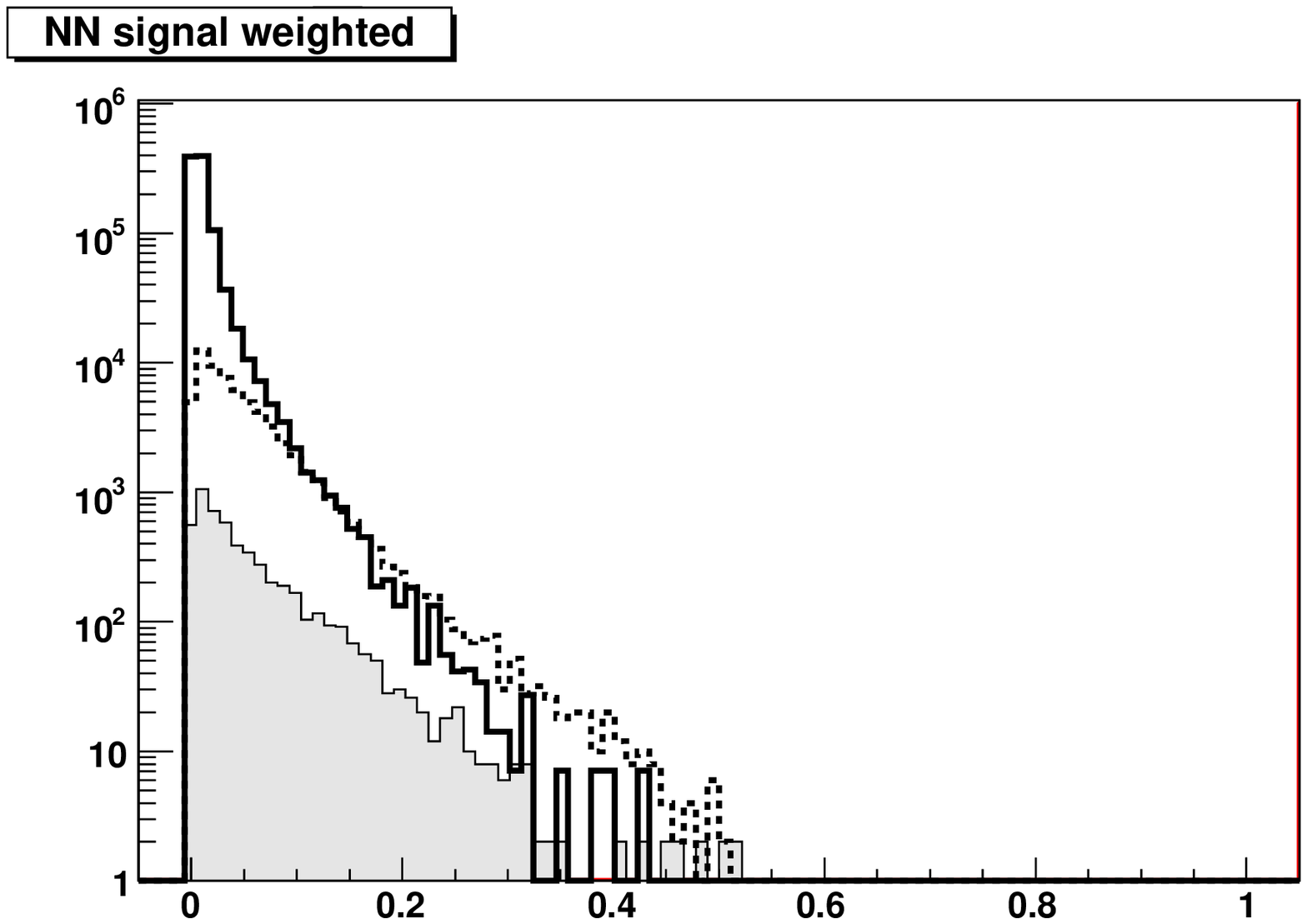}}
\subfloat[]{\includegraphics[scale=0.25]{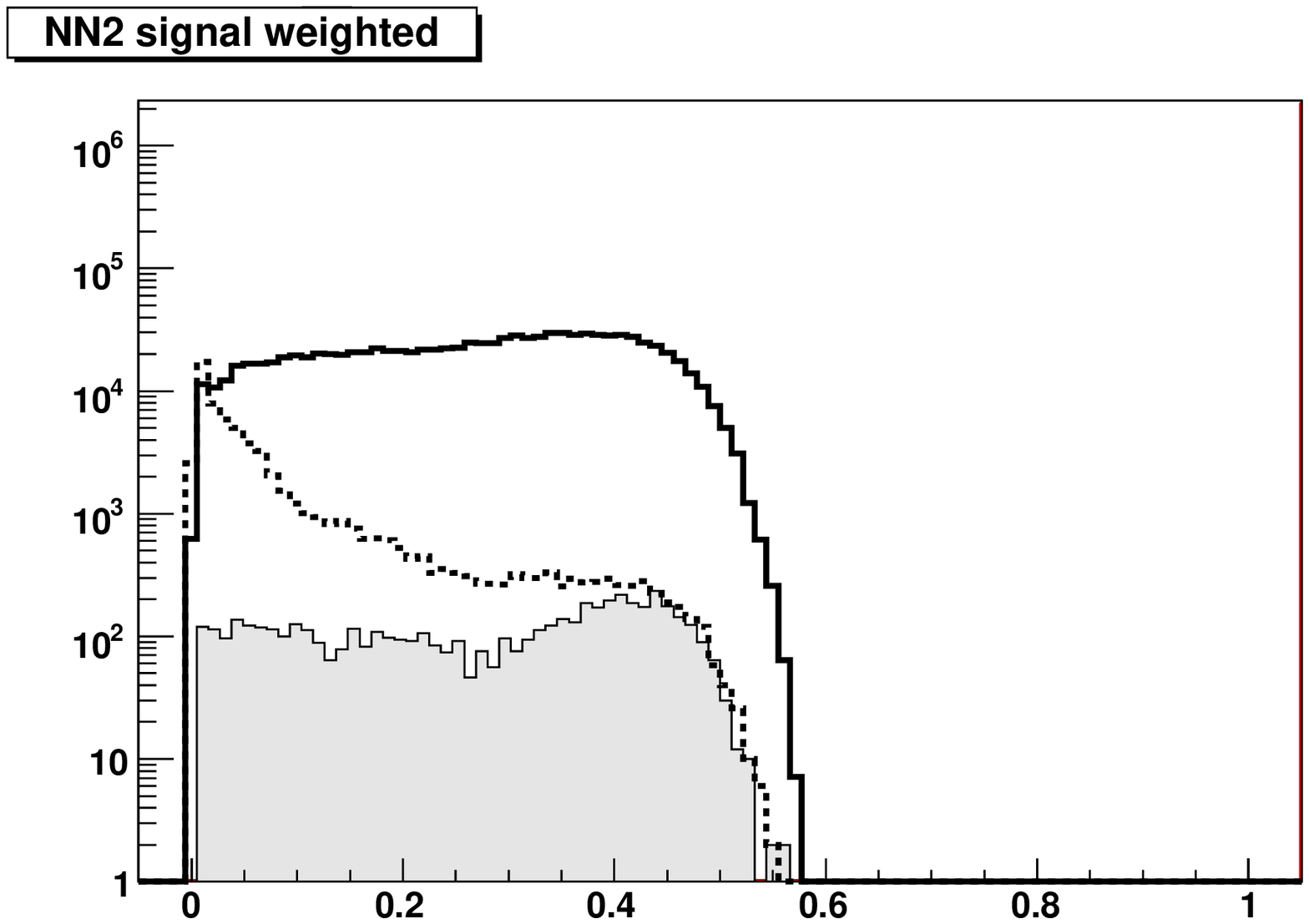}}
\end{center}
\caption{gg: Neutrino channel (a) and (b), and Hadronic channel (c) and (d). First NN corresponds to (a) and (c), and Second NN corresponds to (b) and (d). Solid curves are SM background, dashed curves are inclusive Higgs sample and filled histograms are the signals.}
\label{NNg}
\end{figure}

\begin{figure}[htbp]
\begin{center}
\subfloat[]{\includegraphics[scale=0.35]{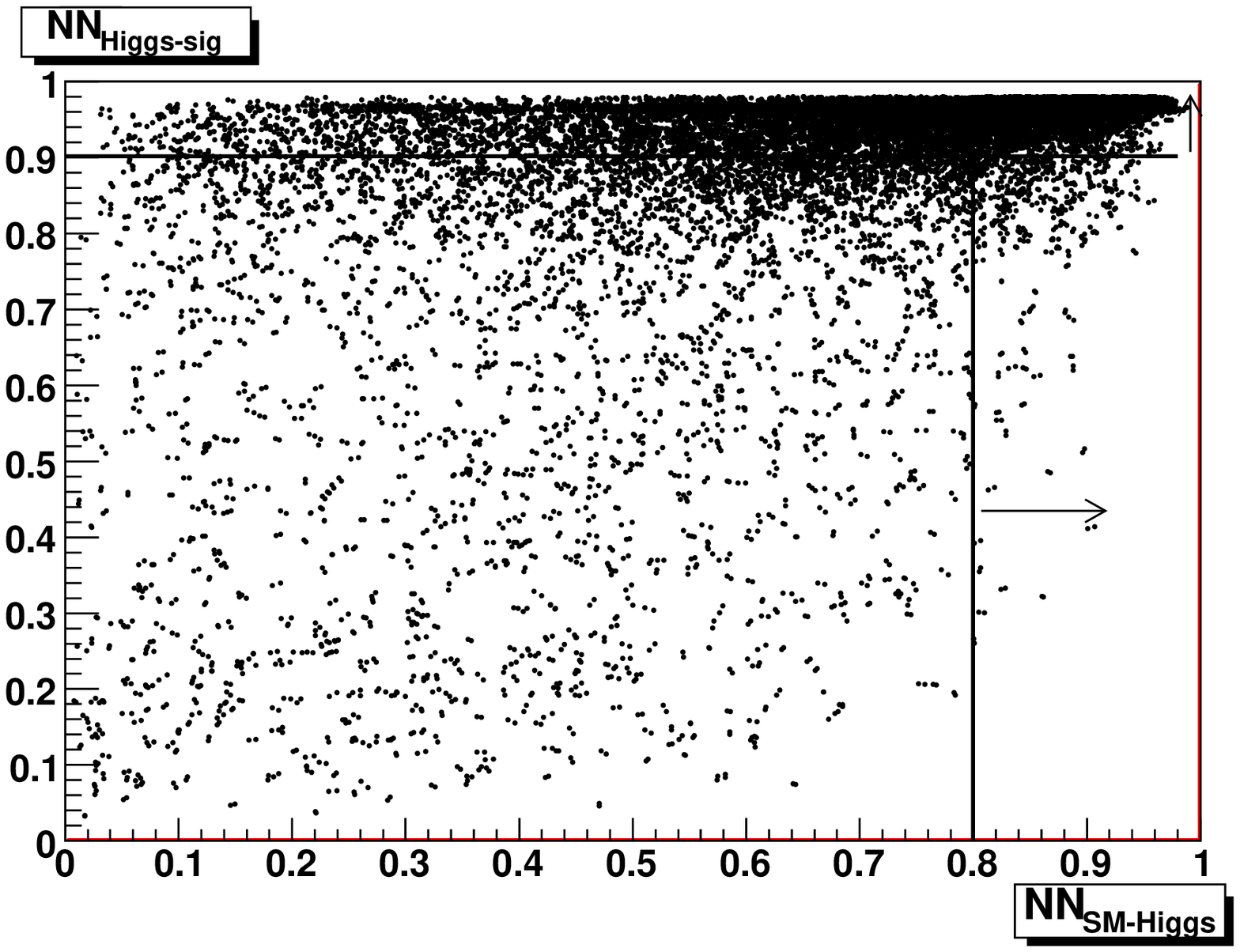}}
\subfloat[]{\includegraphics[scale=0.35]{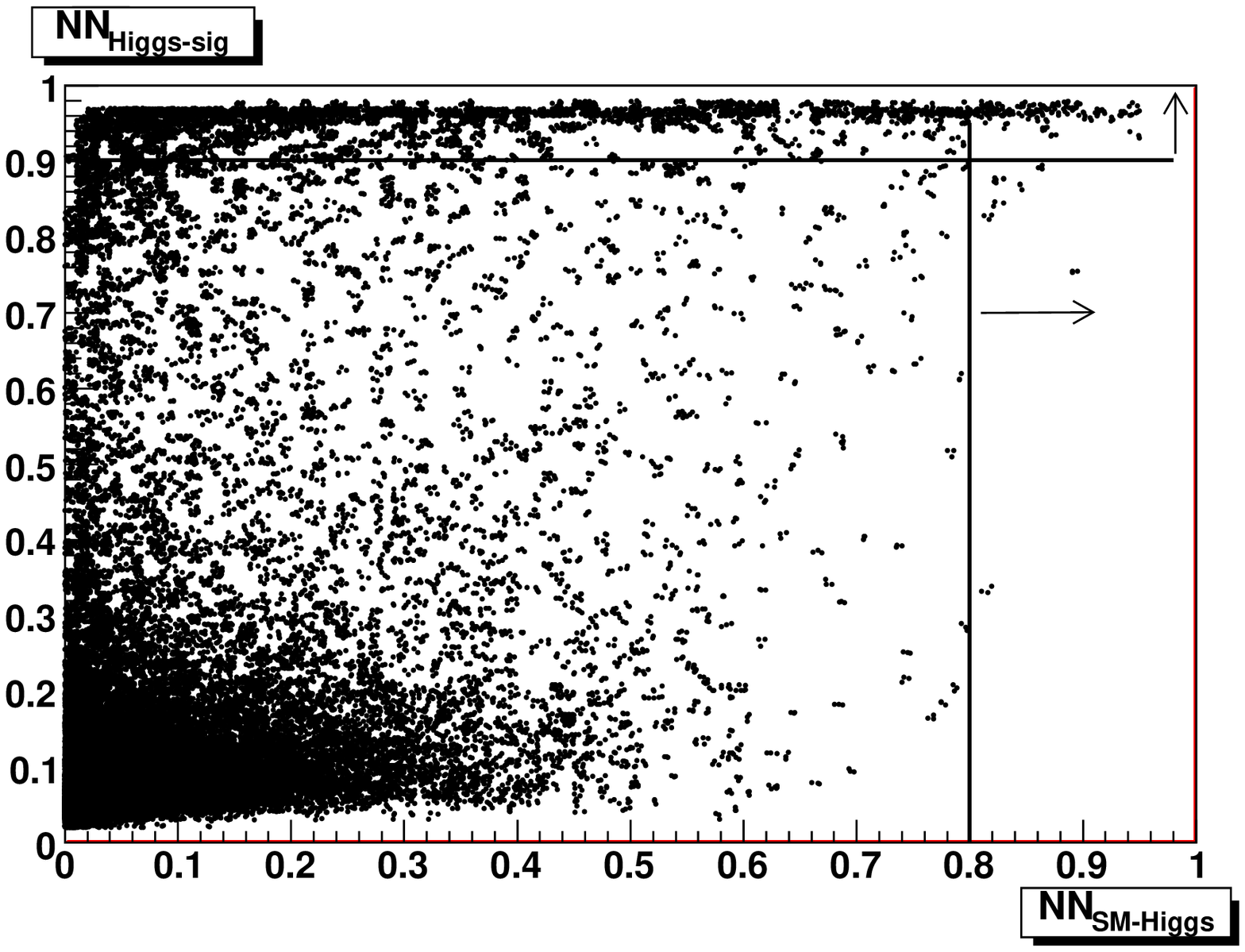}} \\
\subfloat[]{\includegraphics[scale=0.35]{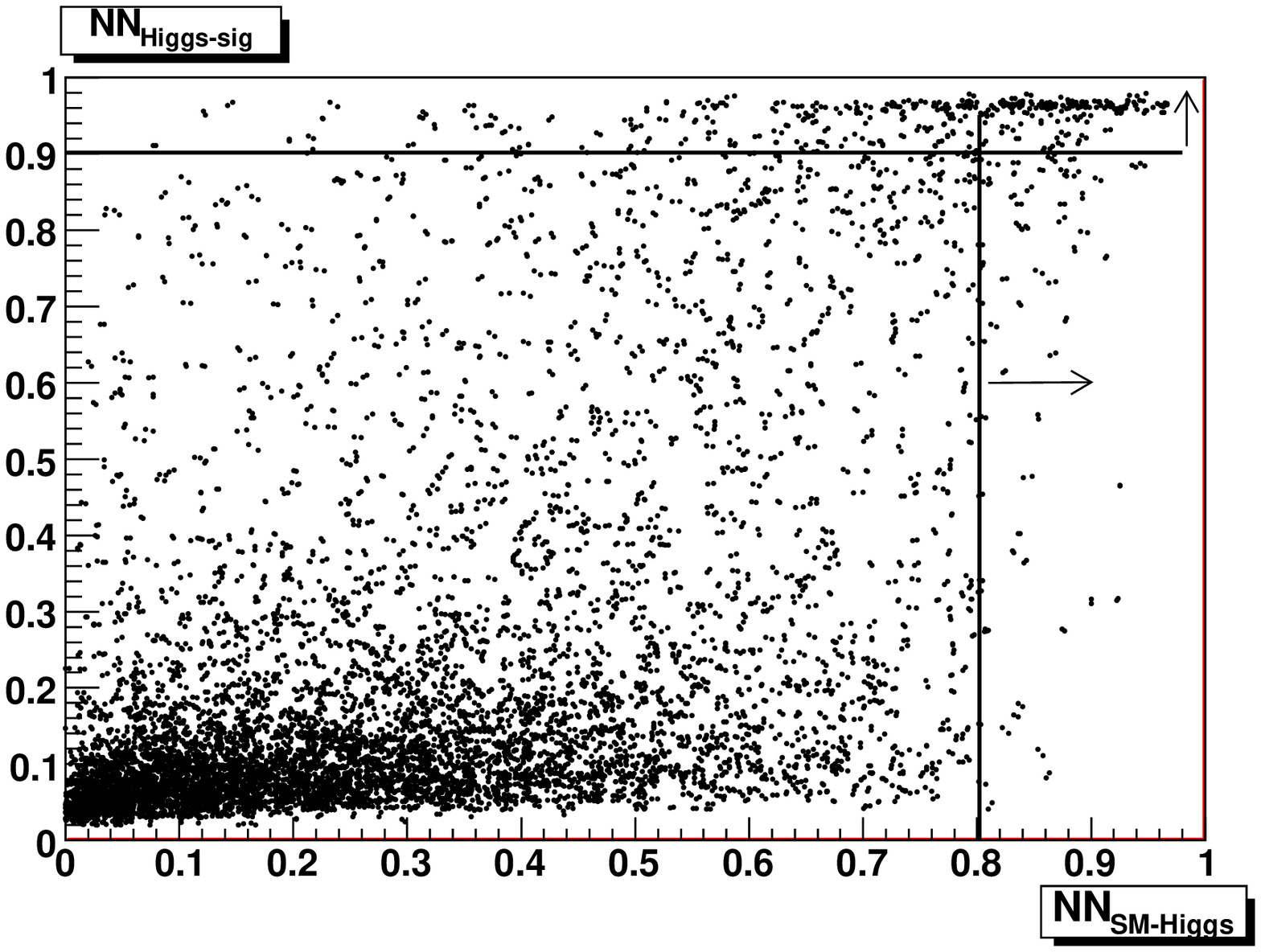}} 
\end{center}
\caption{Neutrino channel $b\bar{b}$: Second NN versus first NN for Signal (a), Standard Model background (b) and Higgs background (c)}
\label{nb1}
\end{figure}

\begin{figure}[htbp]
\begin{center}
\subfloat[]{\includegraphics[scale=0.35]{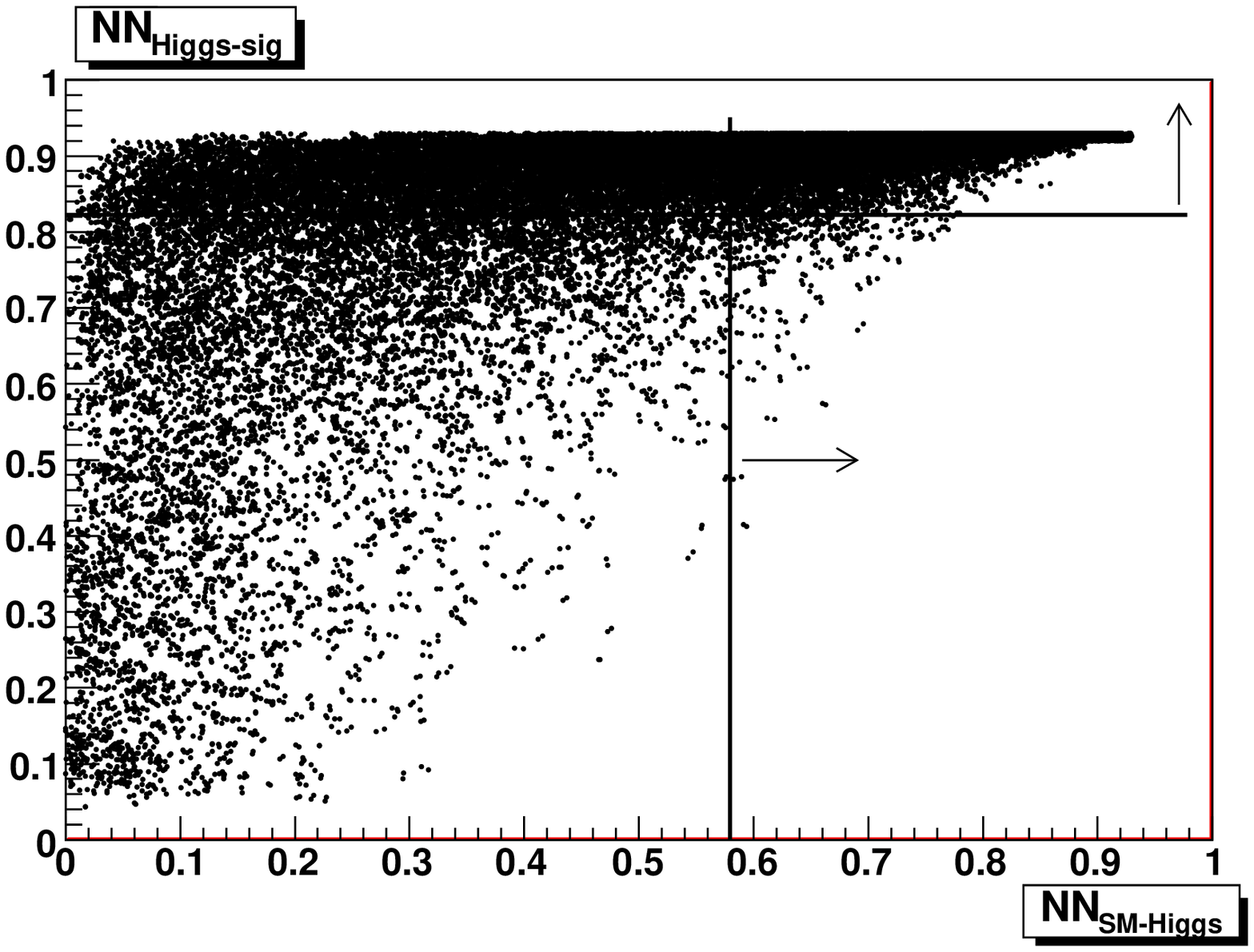}}
\subfloat[]{\includegraphics[scale=0.35]{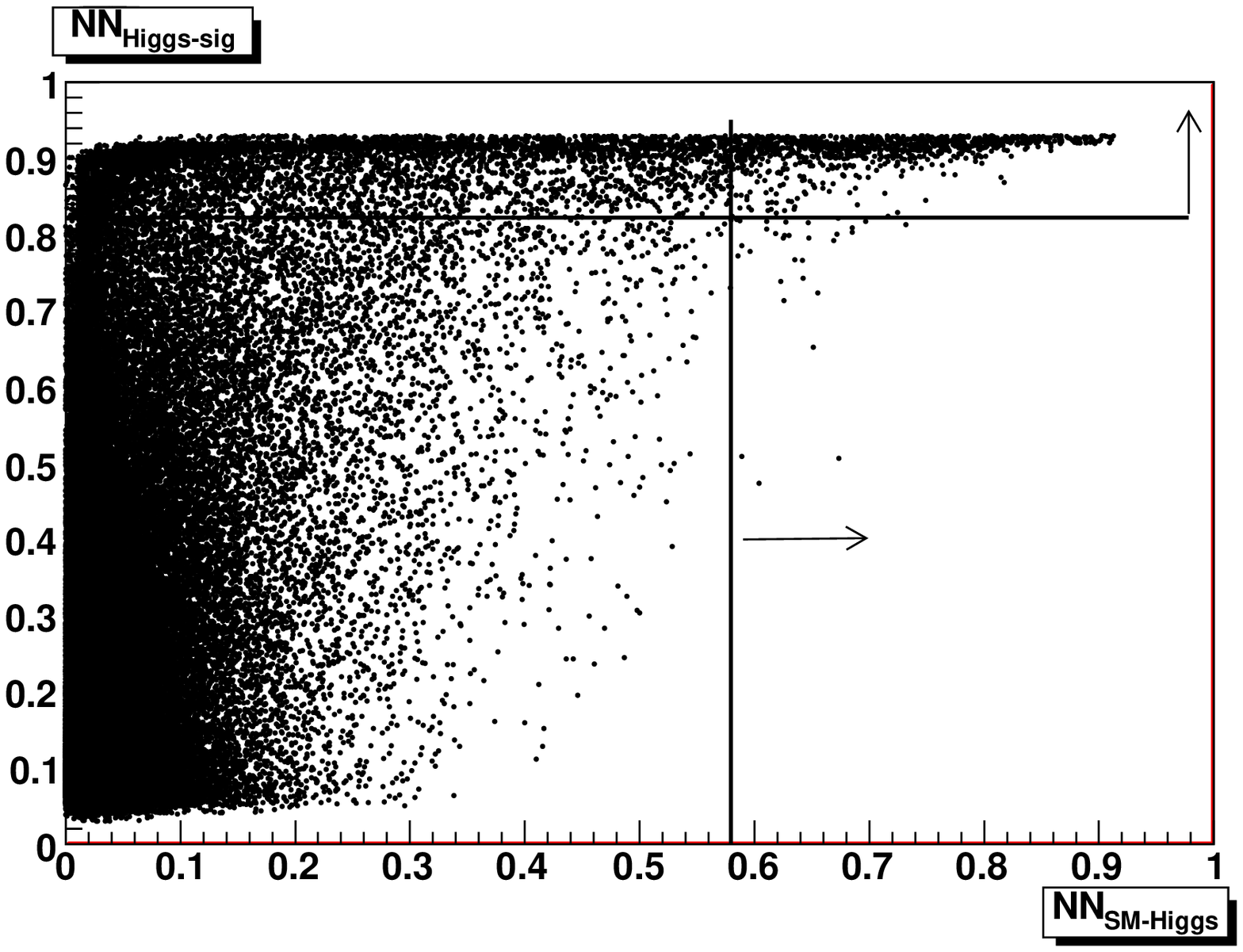}} \\
\subfloat[]{\includegraphics[scale=0.35]{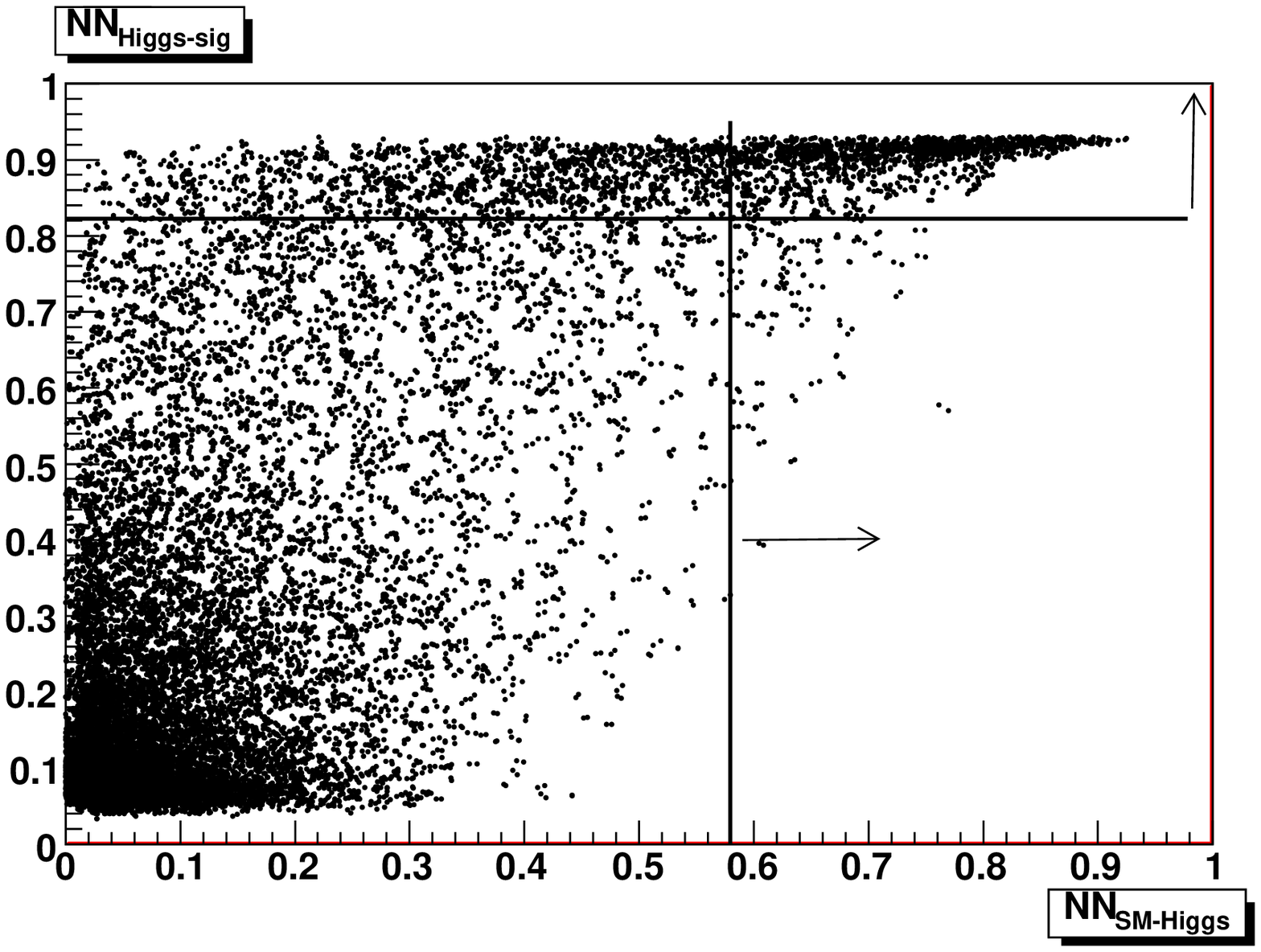}}
\end{center}
\caption{Hadronic channel $b\bar{b}$: Second NN versus first NN for Signal (a), Standard Model background (b) and Higgs background (c)}
\label{nb2}
\end{figure}

\begin{figure}[htbp]
\begin{center}
\subfloat[]{\includegraphics[scale=0.35]{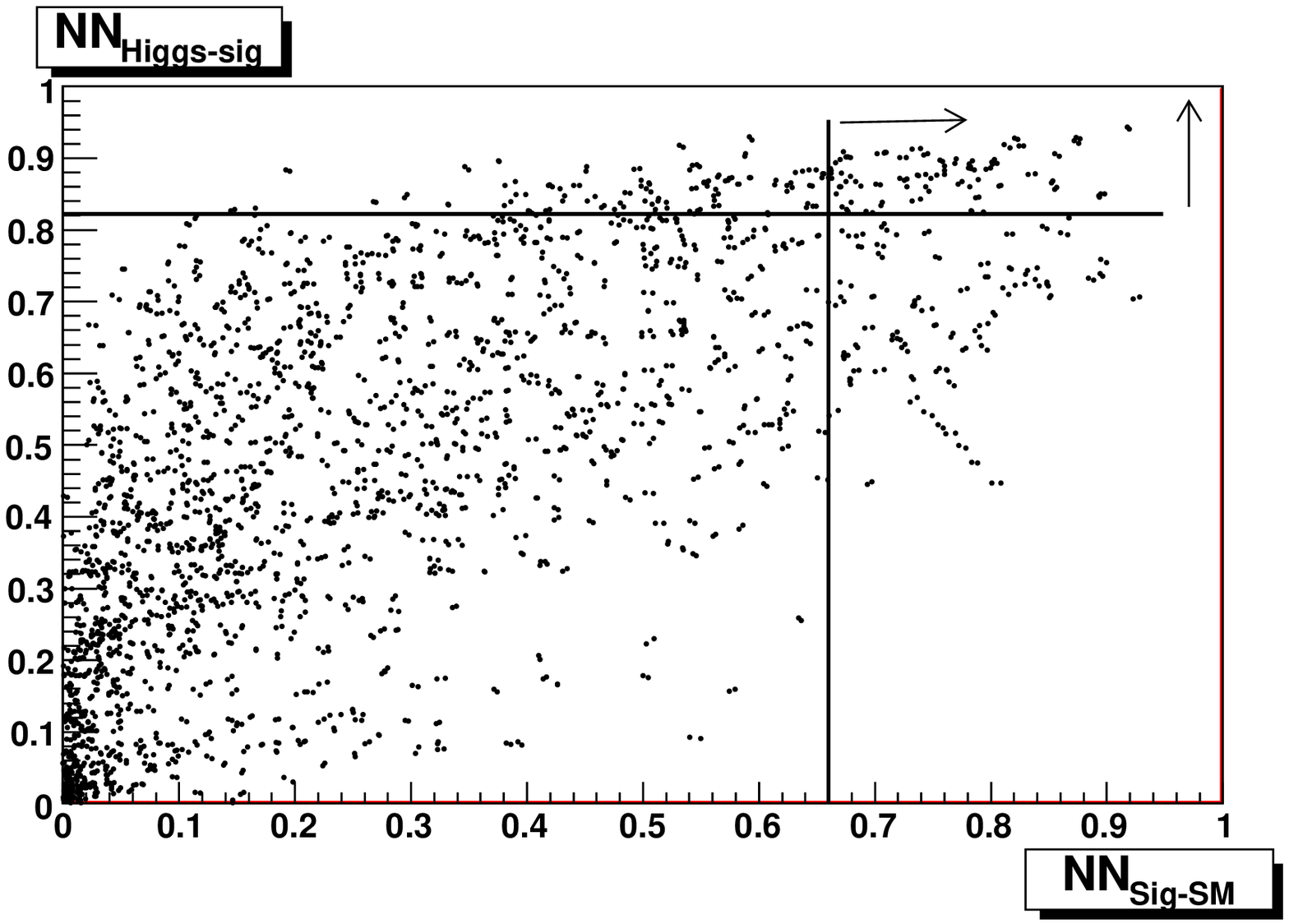}}
\subfloat[]{\includegraphics[scale=0.35]{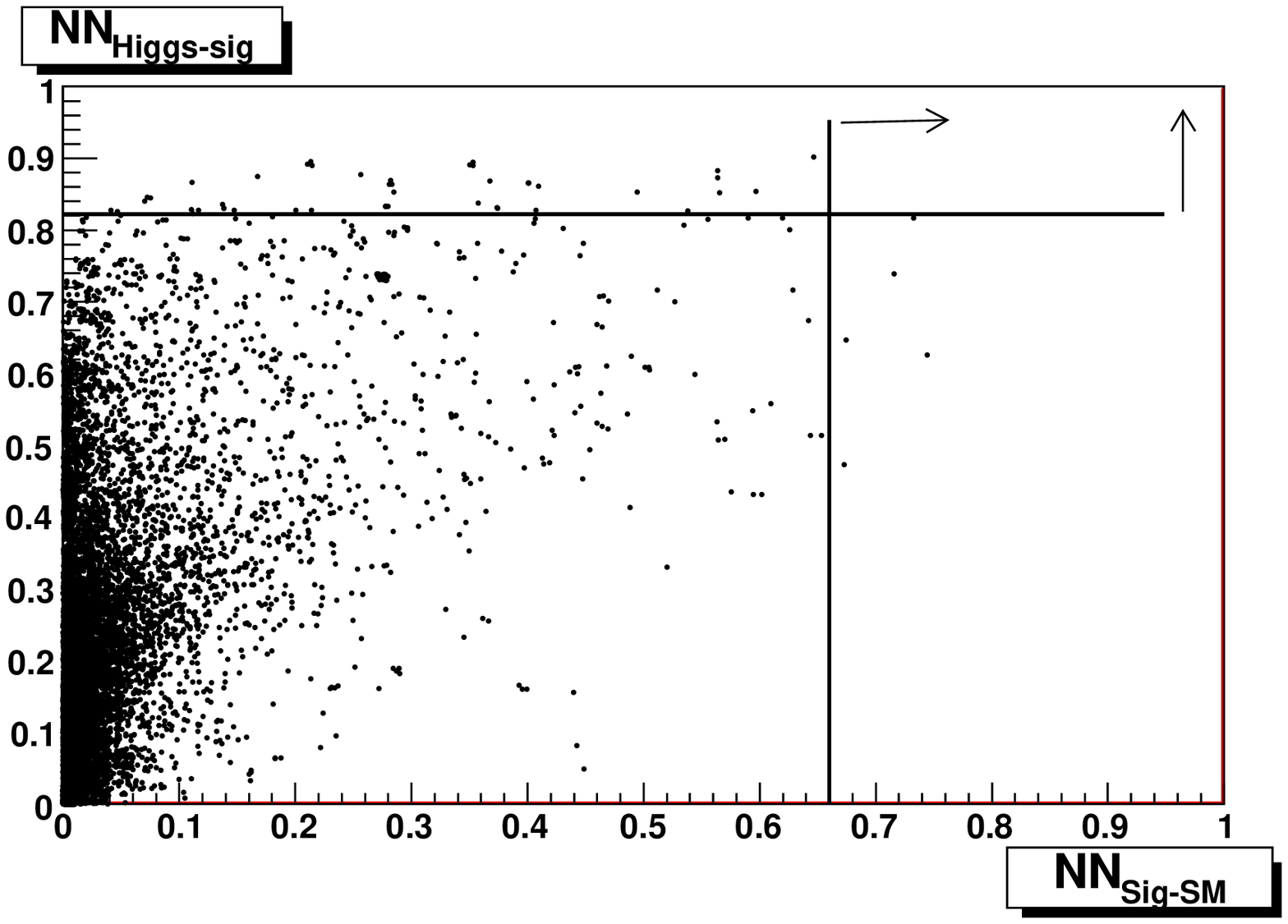}} \\
\subfloat[]{\includegraphics[scale=0.35]{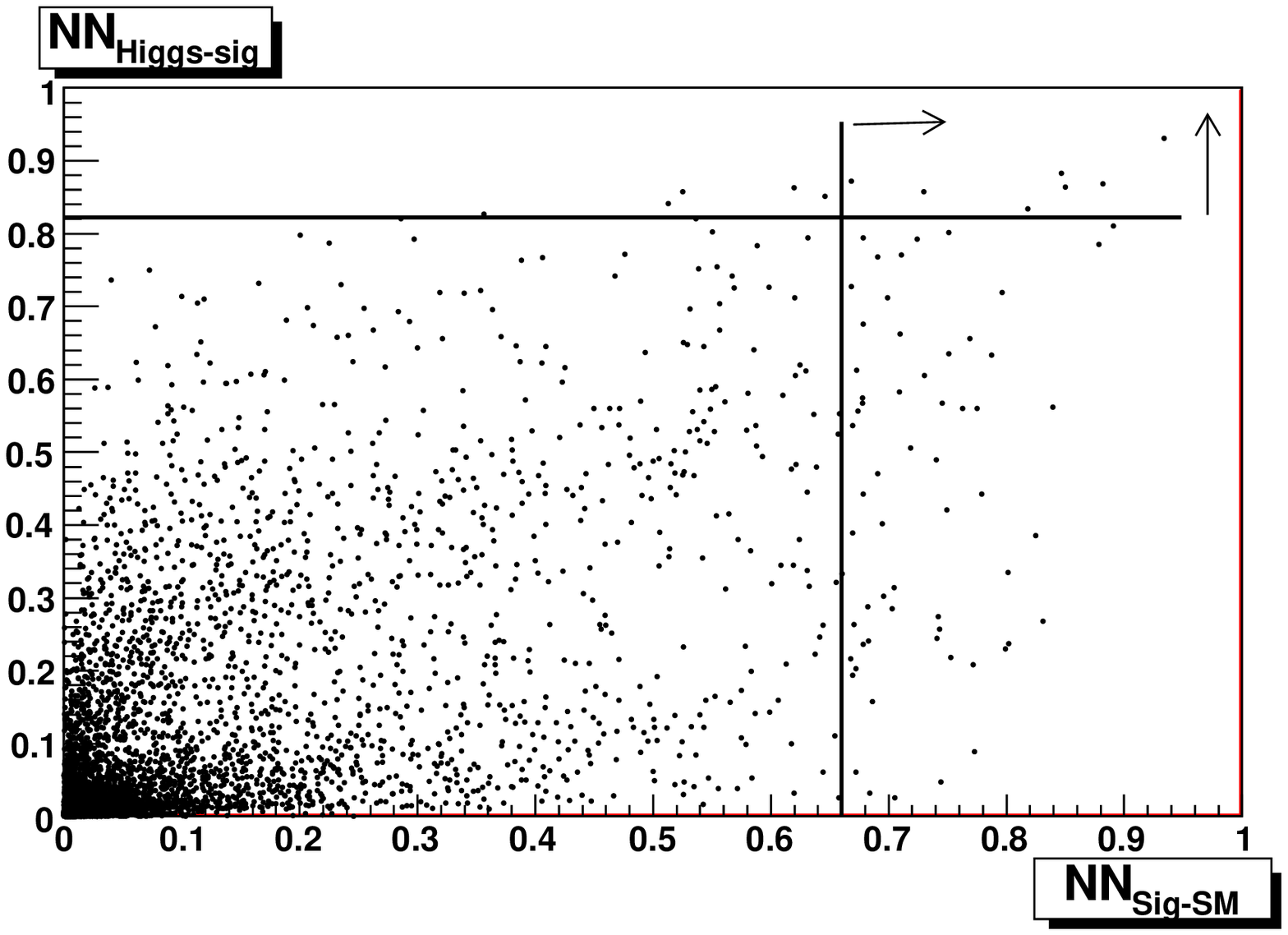}} 
\end{center}
\caption{Neutrino channel gg: Second NN versus first NN for Signal (a), Standard Model background (b) and Higgs background (c)}
\label{ng1}
\end{figure}

\begin{figure}[htbp]
\begin{center}
\subfloat[]{\includegraphics[scale=0.35]{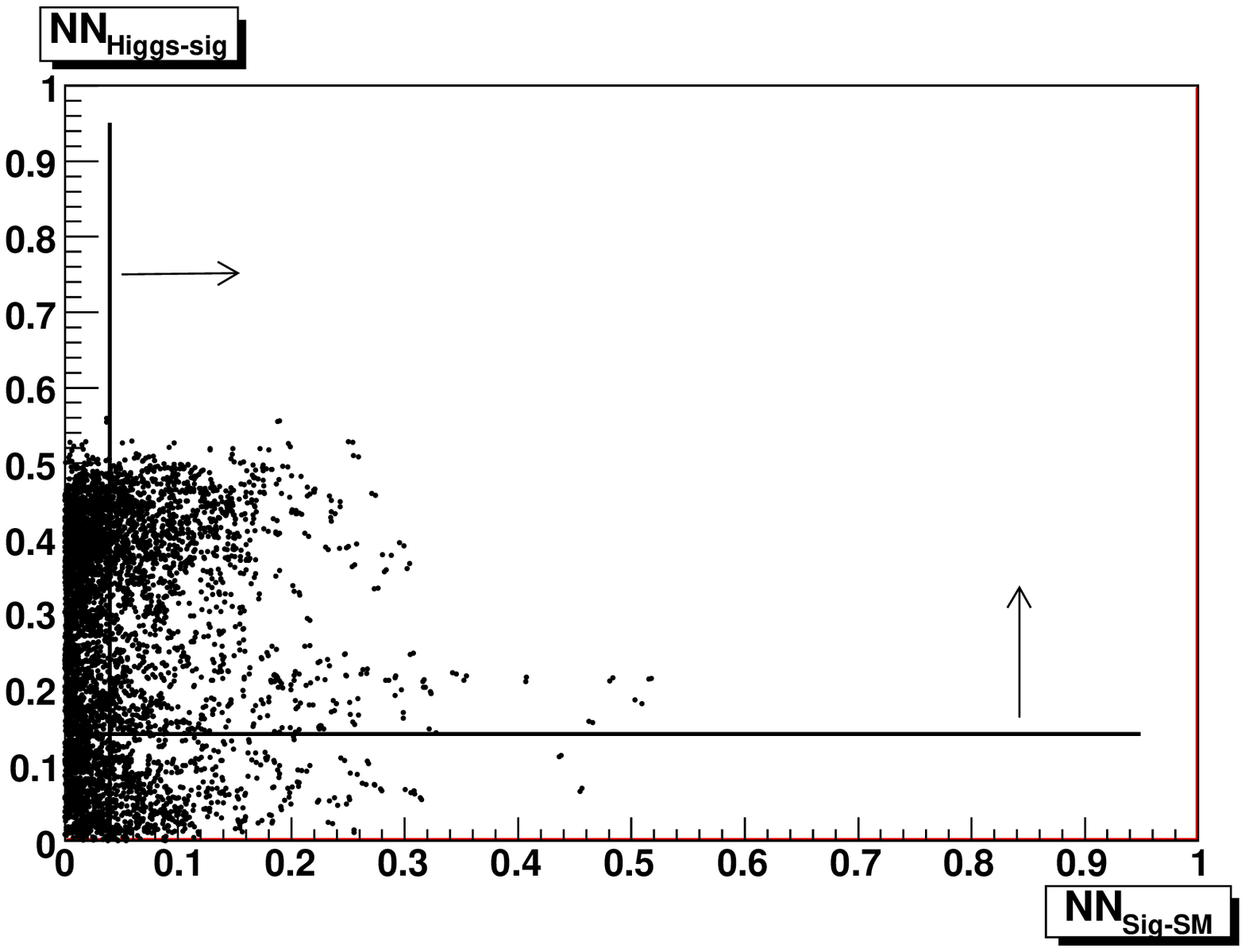}}
\subfloat[]{\includegraphics[scale=0.35]{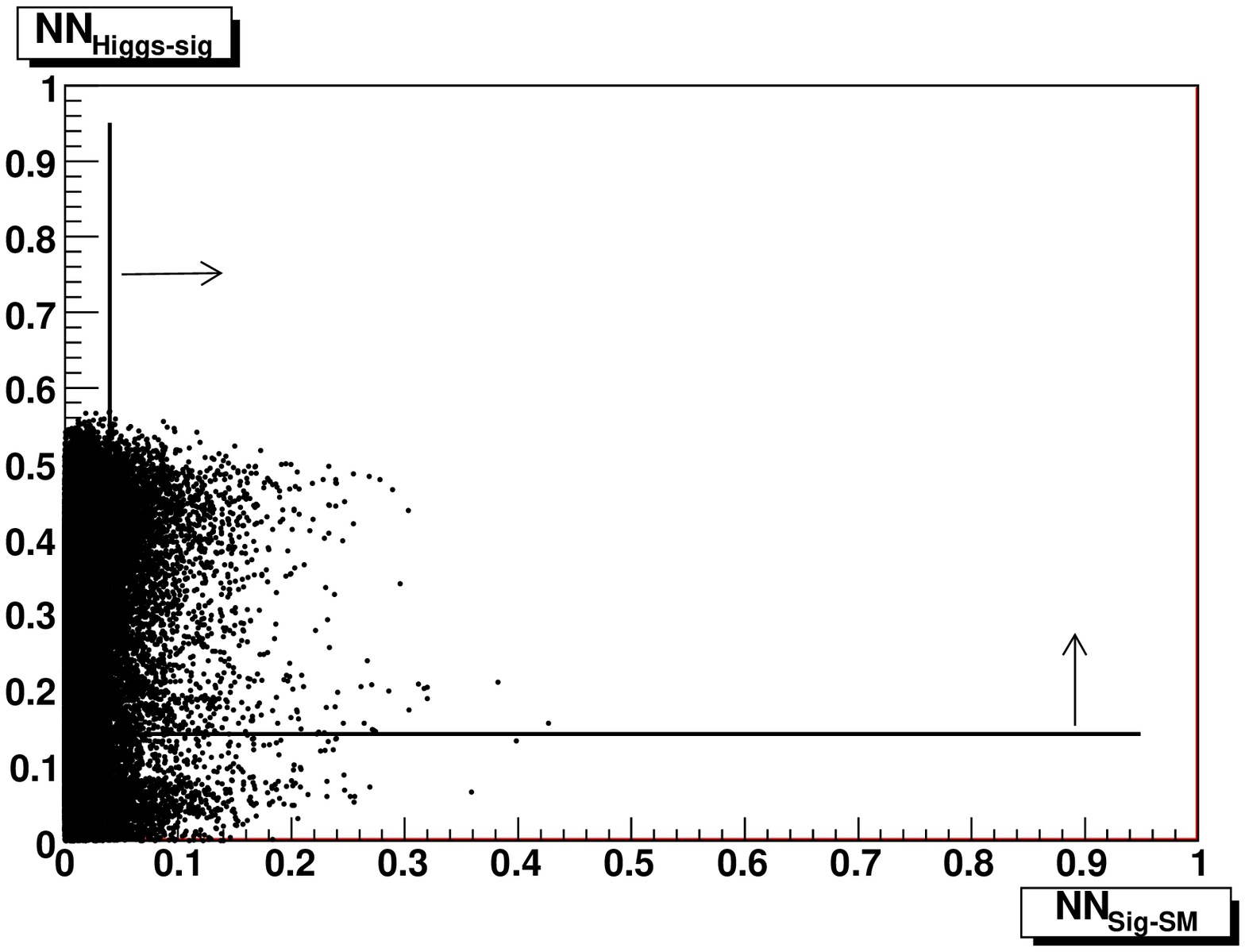}} \\
\subfloat[]{\includegraphics[scale=0.35]{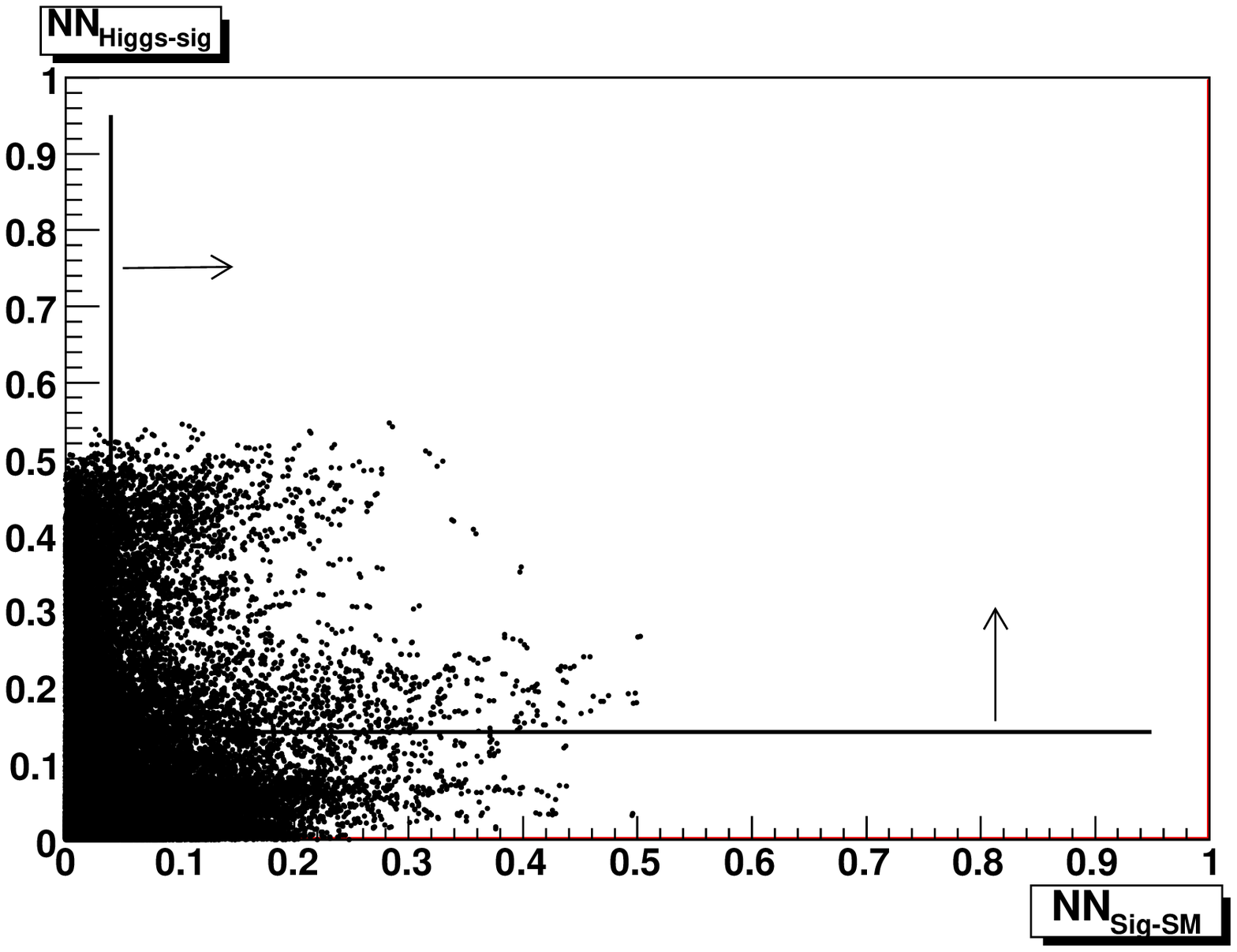}}
\end{center}
\caption{Hadronic channel gg: Second NN versus first NN for Signal (a), Standard Model background (b) and Higgs background (c)}
\label{ng2}
\end{figure}

The events remaining after preselection are categorized using two neural networks implemented in FANN~\cite{fann}. For $b\bar{b}$ the first NN is trained to
distinguish the SM background from the inclusive Higgs sample and to produce the NN$_{SM-Higgs}$ output. In the gg case, the first NN is trained to
distinguish the SM background from the signal sample and to produce the NN$_{Sig-SM}$ output. The second NN is trained to distinguish
the signal from the Higgs background sample and to produce the NN$_{Higgs-signal}$ output. The training is done separately for $b\bar{b}$ and for gg, i.e the training is done twice for the first NN and twice for the second NN. Figures~\ref{NNb} and ~\ref{NNg} show one-dimensional histograms of the trained neutral network outputs. Two-dimensional plots of the outputs of the trained NNs are shown in Figures~\ref{nb1}, ~\ref{nb2}, ~\ref{ng1} and ~\ref{ng2}. Again, all the histograms are normalized to luminosity 250 fb$^{-1}$.

\begin{figure}[htbp]
\begin{center}
\subfloat[]{\includegraphics[scale=0.40]{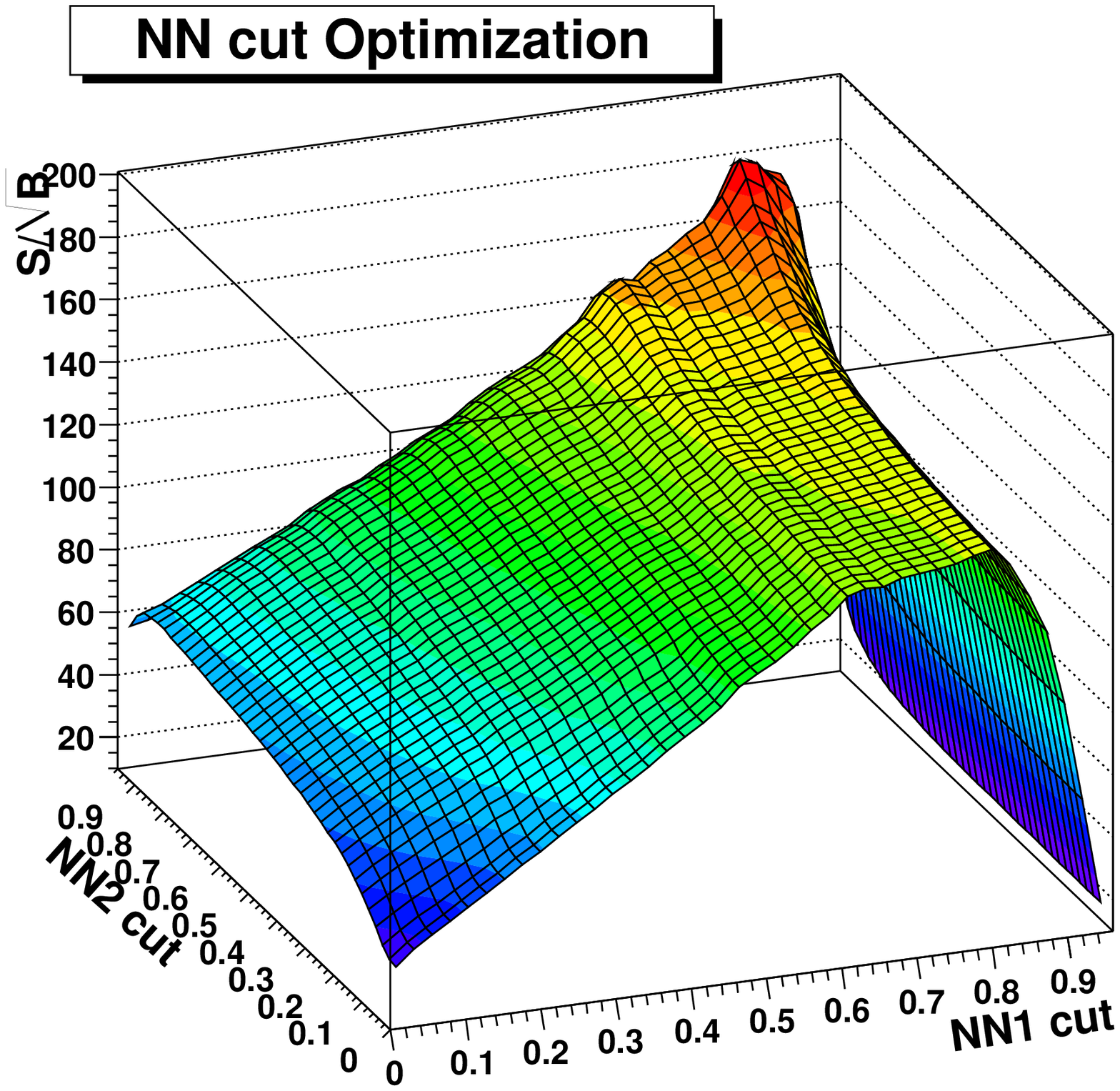}}
\subfloat[]{\includegraphics[scale=0.40]{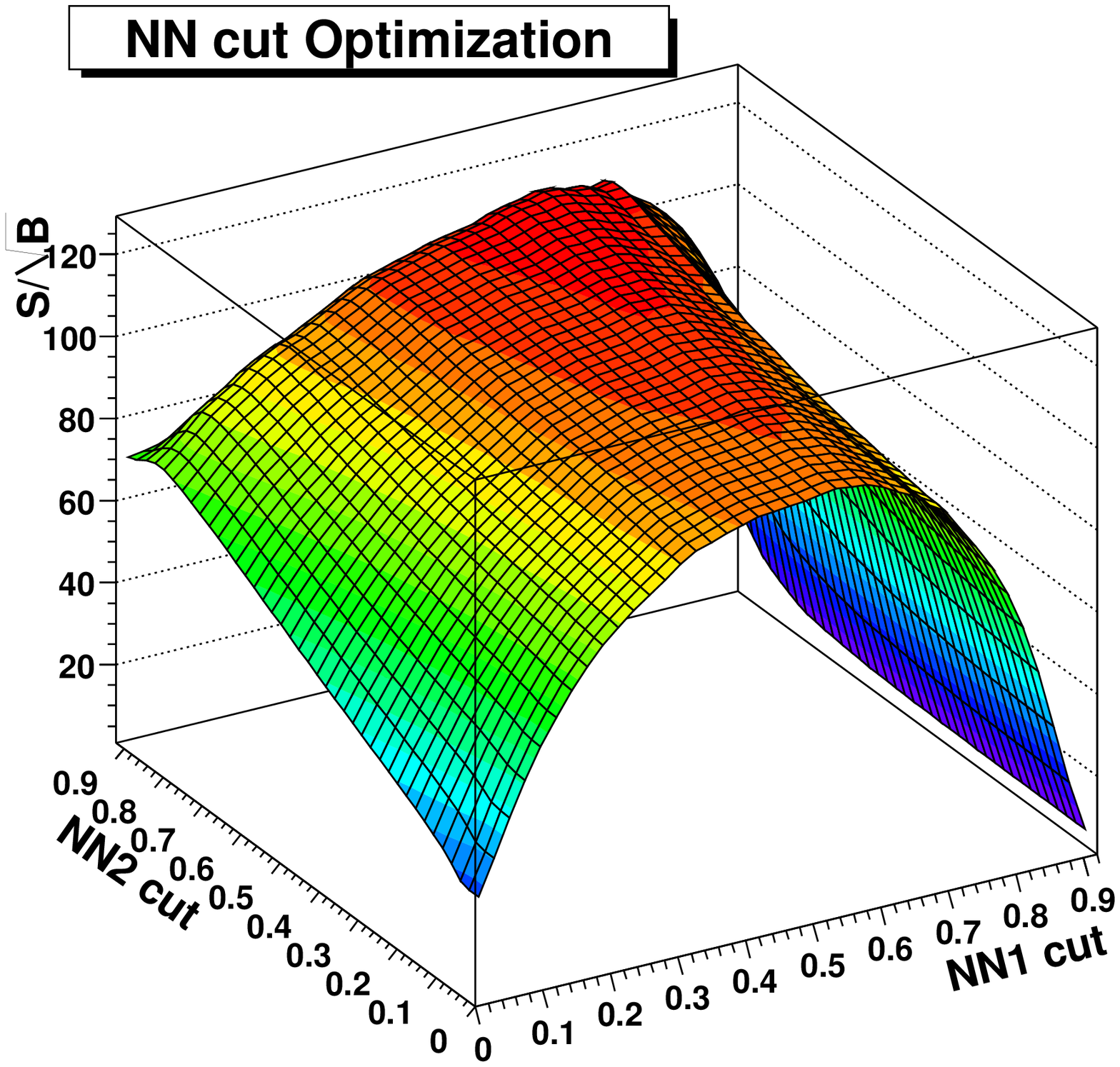}}
\end{center}
\caption{Neural Network Cut optimization for $b\bar{b}$ : neutrino channel (a) and hadronic channel (b)}
\label{nnbcuts}
\end{figure}

\begin{figure}[htbp]
\begin{center}
\subfloat[]{\includegraphics[scale=0.40]{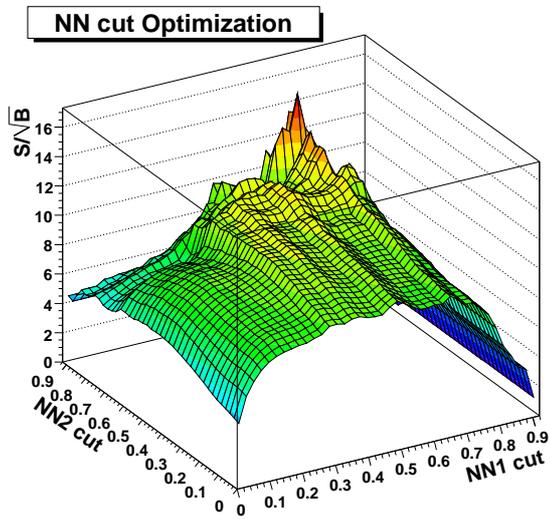}}
\subfloat[]{\includegraphics[scale=0.40]{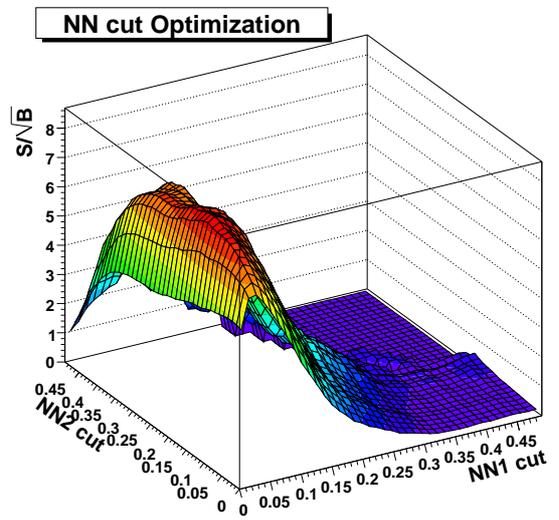}}
\end{center}
\caption{Neural Network Cut optimization for gg : neutrino channel (a) and hadronic channel (b)}
\label{nngcuts}
\end{figure}

The final event samples are obtained after applying cuts on First and Second NNs. These cuts are optimized by choosing values of the NNs that maximise the signal-to-noise ratio as shown in Figures~\ref{nnbcuts} and ~\ref{nngcuts}.

A summary of the results obtained for both the neutrino and hadronic channels is given in Tables~\ref{tab:bbr} and ~\ref{tab:gbr}. 
\begin{table}[h]
\centering
\begin{tabular}{|l|l|l|l|l|}
\hline & Neutrino & Hadronic & Combined \\
\hline
Signal events & 2833 & 8122 & \\ \hline
SM background events & 220 & 4700 & \\ \hline
Higgs background events & 55 & 423 & \\ \hline
Signal efficiency \% & 24.465$\pm$0.004 & 26.213$\pm$0.002 & \\ \hline
Signal $\sigma_{Hbb}$ & 142.7$\pm$2.3 fb & 142.5$\pm$1.9 fb & 142.57$\pm$1.61 fb \\ \hline
Relative uncertainty on $\sigma_{Hbb}$ & 1.9\% & 1.4\% & 1.1\% \\ \hline
\end{tabular}
\caption{Measurement results of H$\rightarrow b\bar{b}$ branching ratio.}
\label{tab:bbr}
\end{table}

\begin{table}[h]
\centering
\begin{tabular}{|l|l|l|l|l|}
\hline & Neutrino & Hadronic & Combined \\
\hline
Signal events & 32 & 524 & \\ \hline
SM background events & 0 & 3621 & \\ \hline
Higgs background events & 4 & 1431 & \\ \hline
Signal efficiency \% & 3.245$\pm$0.006 & 17.673$\pm$0.007 & \\ \hline
Signal $\sigma_{Hgg}$ & 15.1$\pm$1.9 fb & 15.6$\pm$2.6 fb & 15.41$\pm$1.74 fb \\ \hline
Relative uncertainty on $\sigma_{Hgg}$ & 18.7\% & 14.2\% & 11.3\% \\ \hline
\end{tabular}
\caption{Measurement results of H$\rightarrow$gg branching ratio.}
\label{tab:gbr}
\end{table}

Table~\ref{tab:BR} shows the summary of the uncertainties of the Higgs branching ratios to $b\bar{b}$ and gg. Also shown in the table is the uncertainty of the Higgs branching ratio to $c\bar{c}$ as given in~\cite{yb}
\begin{table}[h]
\centering
\begin{tabular}{|l|l|l|}
\hline Channel & $\frac{\Delta BR}{BR}$ \% \\ \hline
H$\rightarrow b\bar{b}$ & 4.8 \\ \hline
H$\rightarrow c\bar{c}$ & 8.4 \\ \hline
H$\rightarrow gg$ & 12.2 \\ \hline
\end{tabular}
\caption{Uncertainties of hadronic Higgs decay branching ratios.}
\label{tab:BR}
\end{table}

\section{Conclusion}

The measurement of the Higgs boson decay branching ratios to bottom quarks and gluons, for a neutral SM Higgs boson of mass 120 GeV, has been studied 
at a centre-of-mass of energy of $\sqrt{s}$ = 250 GeV and a total integrated luminosity of 250$\displaystyle\int$fb$^{-1}$. The analysis is based on full detector simulation and realistic event reconstruction. The uncertainties on the branching ratios are found to be 4.8\% and 12.2\% for bb and gg respectively. The uncertainty in the bb branching ratio is dominated by the uncertainty on the inclusive Higgs-strahlung cross section. A good performance of flavour tagging and the use of neural networks in event selection are critical in obtaining these results.

\bibliographystyle{unsrt}

\end{document}